\documentclass[twocolumn, twocolappendix]{aastex631}

\begin{document}

\title{Absolute Motion of the Infrared Counterpart to Sagittarius A* in the Gaia Celestial Reference Frame 3 and Limits on an Intermediate-mass Black Hole Companion}

\author[0000-0001-7003-0588]{Rebecca A. Lewis-Merrill}
\affiliation{University of California, Los Angeles, Department of Physics and Astronomy, Los Angeles, CA 90095, USA}

\author[0000-0001-9554-6062]{Tuan Do}
\affiliation{University of California, Los Angeles, Department of Physics and Astronomy, Los Angeles, CA 90095, USA}

\author[0000-0003-2874-1196]{Matthew W. Hosek Jr.}
\altaffiliation{Brinson Prize Fellow} 
\affiliation{University of California, Los Angeles, Department of Physics and Astronomy, Los Angeles, CA 90095, USA}

\author{Gregory David Martinez}
\affiliation{University of California, Los Angeles, Department of Physics and Astronomy, Los Angeles, CA 90095, USA}

\author[0000-0001-5972-663X]{Shoko Sakai}
\affiliation{University of California, Los Angeles, Department of Physics and Astronomy, Los Angeles, CA 90095, USA}

\author[0000-0003-2400-7322]{Kelly Kosmo O'Neil}
\affiliation{University of California, Los Angeles, Department of Physics and Astronomy, Los Angeles, CA 90095, USA}
\affiliation{University of Nevada, Reno, Department of Physics, Reno, NV 89557, USA}

\author[0000-0003-4081-1839]{Grant Weldon}
\affiliation{University of California, Los Angeles, Department of Physics and Astronomy, Los Angeles, CA 90095, USA}

\author[0000-0002-2836-117X]{Abhimat Gautam}
\affiliation{University of California, Los Angeles, Department of Physics and Astronomy, Los Angeles, CA 90095, USA}

\author[0009-0004-0026-7757]{Zo\"{e} Haggard}
\affiliation{University of California, Los Angeles, Department of Physics and Astronomy, Los Angeles, CA 90095, USA}

\author[0000-0003-3230-5055]{Andrea M. Ghez}
\affiliation{University of California, Los Angeles, Department of Physics and Astronomy, Los Angeles, CA 90095, USA}

\author[0000-0001-9611-0009]{Jessica R. Lu}
\affiliation{University of California, Berkeley, Astronomy Department, Berkeley, CA 94720, USA}

\author{Keith Matthews}
\affiliation{Division of Physics, Mathematics and Astronomy, California Institute of Technology, Pasadena, CA 91125, USA}

\begin{abstract}
    We report the first proper motion and acceleration measurements of the infrared (IR) counterpart to Sagittarius A* (Sgr A*-IR), the supermassive black hole (SMBH) at the center of our Galaxy, in the Gaia-Celestial Reference Frame (Gaia-CRF3). This reference frame realizes the International Celestial Reference System (ICRS), which is an absolute reference coordinate system defined by quasars. A combination of Gaia and Hubble Space Telescope data was used to transform Keck adaptive optics (AO) observations into Gaia-CRF3. We developed a method for selecting reference stars that minimizes astrometric transformation errors (statistical error = $0.10-0.63$ mas) and drift of the coordinate system (systematic error $\sim 0.01$ mas/yr). We find the proper motion of Sgr A*-IR in Gaia-CRF3 to be $\mu_{\alpha^{*}} = -3.093 \pm 0.085$ mas yr$^{-1}$ and $\mu_{\delta}= -5.62 \pm 0.13$ mas yr$^{-1}$ with the initial position at $t_{0} = 2016.0$ of R.A. = 266.41680848 $\pm$ 0.00000029 deg and DEC = -29.00783947 $\pm$ 0.00000050 deg, which translates to a precision of 1.05 mas in R.A. and 1.79 mas DEC. This is consistent with the astrometric measurements of the radio counterpart to Sgr A* by \citet{Xu_2022}. We also place a $2\sigma$ upper constraint of the acceleration of Sgr A*-IR on the sky at 0.061 mas yr$^{-2}$. This acceleration limit on Sgr A*-IR excludes any intermediate-mass black hole companion with mass $\gtrsim 4\times 10^{4}$ $M_{\odot}$ within a distance of $\sim$0.01 pc, consistent with previous studies. With the release of Gaia Data Release 4, we predict these limits will be improved by at least a factor of two.
\end{abstract}


\section{Introduction}

The Milky Way's Galactic center (GC) allows us to study a supermassive black hole (SMBH) and its surrounding environment at a level of detail not possible elsewhere in the Universe. Due to its proximity, we can obtain high precision astrometry of the emissive source associated with the SMBH at the GC, called Sagittarius A* (Sgr A*), as well as of the stars in orbit around it \citep{Schodel_2002,Eckart_2002,Ghez_2003}. Such measurements can reveal the gravitational influence of massive objects on the SMBH, for example, an intermediate-mass black hole (IMBH) companion \citep[e.g.][]{Hansen_2003, Reid_2004, Naoz_2020, Abuter_2020, Will_2023, GRAVITY_2023}. IMBHs exhibit masses between $\sim$10$^{2}$ $M_{\odot}$ and $\sim$10$^{5}$ $M_{\odot}$. This range is more massive than the stellar-mass black holes formed from massive stars via stellar evolution \citep[M $\lesssim$ 100 $M_{\odot}$; e.g.,][]{Madau_2001, Belczynski_2010}, but less massive than most SMBHs found in galactic nuclei \citep[M $\gtrsim 10^5$ $M_{\odot}$; e.g.][]{Kormendy_2013}. IMBHs are predicted to congregate in the deep potential wells of galactic centers via multiple proposed formation mechanisms \citep[e.g.][]{Rashkov_2014,Rose_2022} and may seed the formation of SMBHs in the early Universe \citep[e.g.][]{Greene_2020}. In addition, IMBHs have been invoked to explain the formation of the young stars orbiting Sgr A* \citep{Hansen_2003}. However, there is much debate about the existence of IMBHs, and proposed candidates at the GC have been difficult to confirm \citep[e.g.][]{Schodel_2005,Tsuboi_2017, Tsuboi_2020, Takekawa_2019, Takekawa_2020, Zhu_2020, Hosseini_2024, Roychowdhury_2025}.

There are several avenues to dynamically constrain the properties of a potential IMBH companion to the SMBH associated with Sgr A*. One approach is to consider a three-body system between the SMBH, the star S0-2, and the hypothetical IMBH. Changes in the orbital parameters of S0-2 over time can be used to constrain the parameters of a potential companion interior and exterior to the orbit of S0-2 \citep{Gualandris_2010,Abuter_2020, Will_2023, GRAVITY_2023}. Along a similar line of three-body interactions, constraints on an IMBH companion to the SMBH can also be made using the distribution of hypervelocity stars originating from the GC. This constraint depends on dynamical interactions with the SMBH-IMBH binary ejecting single stars with a high enough velocity to escape the Galaxy \citep{Evans_2023}. Finally, an IMBH companion can also be constrained by the presence (or absence) of acceleration in the observed motion of Sgr A* \citep{Reid_2004}. In the case of no significant sources of mass near Sgr A*, the SMBH is expected to be at rest with respect to the dynamical center of the galaxy. Thus, the observed motion of Sgr A* from Earth is expected to be linear in its sky motion, dominated by the reflex motion of the Sun's orbit in the Galaxy. In the presence of an IMBH companion, Sgr A* will exhibit orbital motion around the barycenter of the IMBH-SMBH system, which would appear as an apparent acceleration (or oscillations) on top of its linear motion as observed from Earth. For example, radio observations can measure the proper motion and constrain the acceleration on the sky of Sgr A* \citep[Sgr A*-radio;][]{Reid_2004,Reid_2020,Xu_2022}. So far, these methods have not detected an IMBH, but have placed limits on the location and mass of a hidden companion. 

It is also possible to also measure the reflex motion of Sgr A* in the infrared (IR) by measuring the motion of the infrared counterpart to Sgr A* (Sgr A*-IR). Its IR emission has been studied extensively in adaptive optics \citep[AO; e.g., ][]{Genzel_2003, Do_2009, Dodds-Eden_2011, Witzel_2018, Weldon_2023}. The measurement of the motion of Sgr A*-IR can provide an independent verification of the radio results, and benefits from the large set of archival and ongoing IR imaging datasets monitoring stellar dynamics at the GC \citep[e.g.][]{Ghez_Jan2005,Do_2019,Abuter_2020}.

However, measuring the absolute motion of Sgr A*-IR has not yet been possible due to the lack of an appropriate reference frame. The reference frames typically used for AO IR observations are constructed such that the Sgr A*-radio source is at rest. Thus, astrometric measurements in these frames are insensitive to the motion of Sgr A* itself \citep{Ghez_2008,Gillessen_2009,Yelda_2014,Plewa_2015, Sakai_2019}. We refer to these as relative reference frames. Measuring the motion of Sgr A*-IR requires a reference frame defined independently from Sgr A*, for example with respect to extragalactic sources. We label such a reference frame as an absolute reference frame. Examples of absolute reference frames include the radio-based International Celestial Reference Frame \citep[ICRF;][]{Charlot_2020} and the optical-based Gaia Celestial Reference Frame 3 \citep[Gaia-CRF3;][]{Klioner_2022}. Gaia measurements of stars offer a way to place observations of Sgr A*-IR and its surrounding region into Gaia-CRF3. However, there are no Gaia sources with reliable astrometry within the field of view (FOV) of typical AO observations \citep[R$\lesssim$10";][]{Ghez_2008}, where we define reliability by the quality cuts applied to reference sources in \citet{Hosek_2025}.

We can overcome this issue by using Hubble Space Telescope (HST) Wide Field Camera 3 (WFC3) IR observations of the GC. HST-WFC3 has a wider field of view than AO by about 144 times, and so there are many common stars between the Gaia Data Release 3 (DR3) and HST star lists. This overlap allowed for the creation of the HST-Gaia catalog of astrometric reference stars by \citet{Hosek_2025}. Since the HST observations are at IR wavelengths, there is also significant overlap between the sources observable with HST and with ground AO. This provides sources with known astrometry in Gaia-CRF3 (via the HST catalog) which can be used as reference stars to define the reference frame for the ground AO observations. 

In this paper, we make the first measurement of the absolute proper motion of Sgr A*-IR in Gaia-CRF3. We also place corresponding constraints on an IMBH companion. The paper is structured as follows. Section \ref{sec:data} summarizes the AO datasets of the GC in the near-IR (NIR) and the HST-Gaia catalog reference sources that are used in this study. Section \ref{sec:methods} shows our methods for transforming the observations into Gaia-CRF3, for selecting Sgr A*-IR astrometric points, and for fitting the proper motion of Sgr A*-IR. Section \ref{sec:results} presents our results for the proper motion of Sgr A*-IR, including constraints on its acceleration on the sky. Section \ref{sec:discussion} shows the resulting limits on the existence of an IMBH companion, compares our limits against previous studies, and discusses how our measurement of the motion of Sgr A*-IR can be improved with further Gaia data. We give our conclusions in section \ref{sec:conclusion}.

\section{Observations and Datasets}
\label{sec:data}

We use two datasets as input to this study. The first is from the GC Orbits Initiative (GCOI), which is a long-term program at the W. M. Keck Observatory designed to measure the orbits of stars around the SMBH associated with Sgr A* (PI: A. Ghez) and which contains the NIR observations of Sgr A* used in this study (Section \ref{subsec:keck_ao_obs}). The second is the GC HST-Gaia catalog from \citet{Hosek_2025}, which provides kinematic models for potential astrometric reference stars for aligning the GCOI astrometric measurements to Gaia-CRF3 (Section \ref{subsec:hst_gaia_catalog}).

\subsection{Keck Adaptive Optics Observations}
\label{subsec:keck_ao_obs}

\begin{figure*}
    \centering
    \includegraphics[width=1\linewidth]{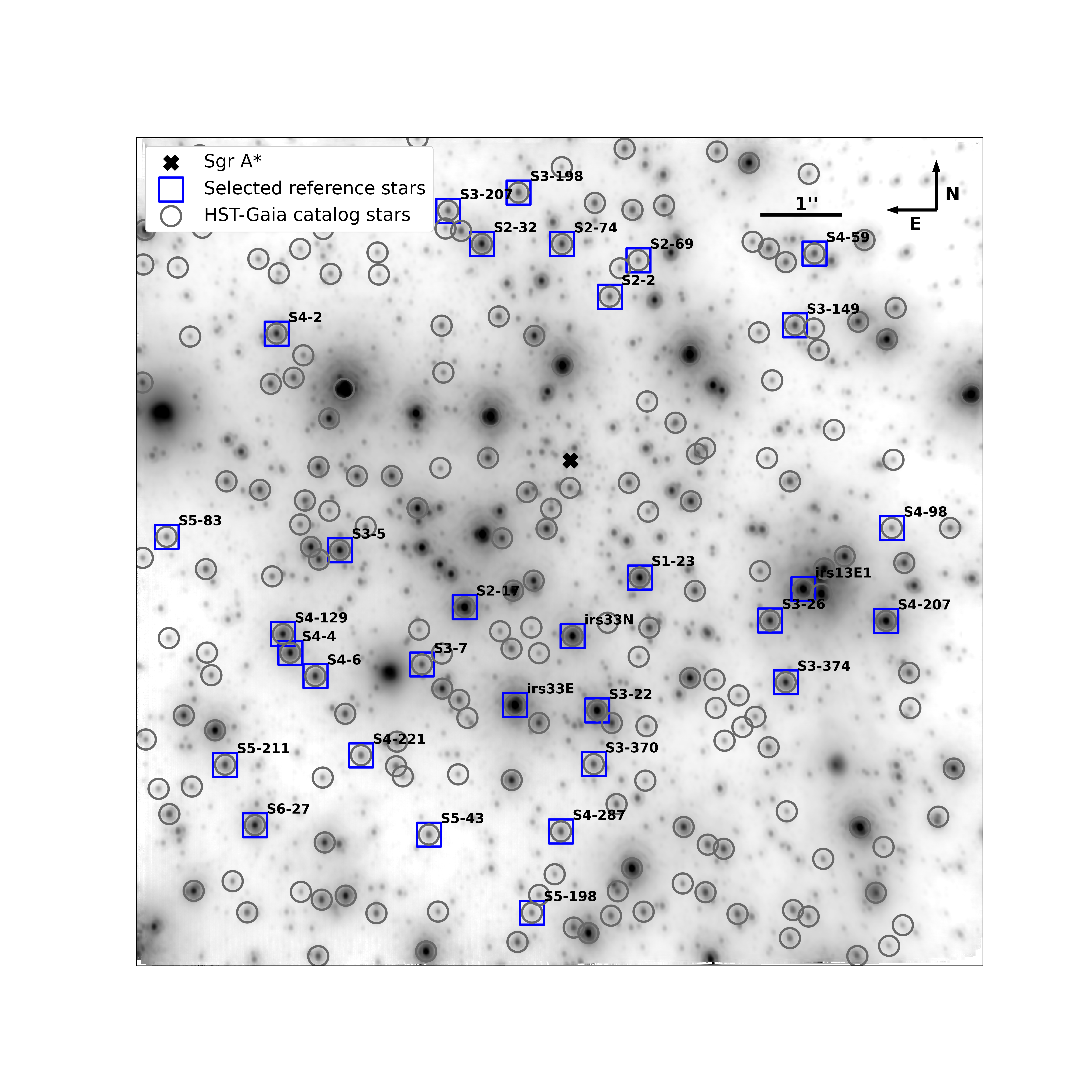}
    \caption{Keck AO IR image of our field (the central 10" of the GC) at observation epoch 2012 July 24. Gray circles represent the positions of HST-Gaia catalog stars, and in addition, final reference stars are marked by a blue square. Sgr A*-IR is marked in the middle with a black cross.}
    \label{fig:selected_ref_stars}
\end{figure*}

We use a subset of GCOI data originally published by \citet{ONeil_2023} and references therein for this work. The observations we use were obtained with the 10 m W.M. Keck II telescope between 2006 May and 2023 May with the Keck II laser guide star AO (LGSAO) system \citep{van_Dam_2006, Wizinowich_2006} and with the NIR camera NIRC2 (PI: K. Matthews) (see Table \ref{tab:sgra_nir_tab}). We also restrict images we use to those taken in the \textit{K}$^{\prime}$ bandpass filter ($\lambda_{0} = 2.212 \,\mu\mathrm{m}$, $\delta\lambda_{0} = 0.35 \,\mu\mathrm{m}$). We refer to these \textit{K}$^{\prime}$-band AO observations as the Keck AO images (Figure \ref{fig:selected_ref_stars}). Details of the observations, their data reduction, and the output Keck AO star lists can be found in \citet{ONeil_2023} and references therein. The Keck AO star lists contain positional and K$^\prime$ photometric measurements of stars in the central 10" of the GC and their associated uncertainties. For calculation of the uncertainties, the frames used to construct the final images for each observation epoch are divided into three subsets with similar Strehl and FWHM statistics. The frames of each subset of a given epoch are then averaged, while being weighted by Strehl ratio, to create three submaps for that epoch.

\begin{figure}
    \centering
    \includegraphics[width=1\linewidth]{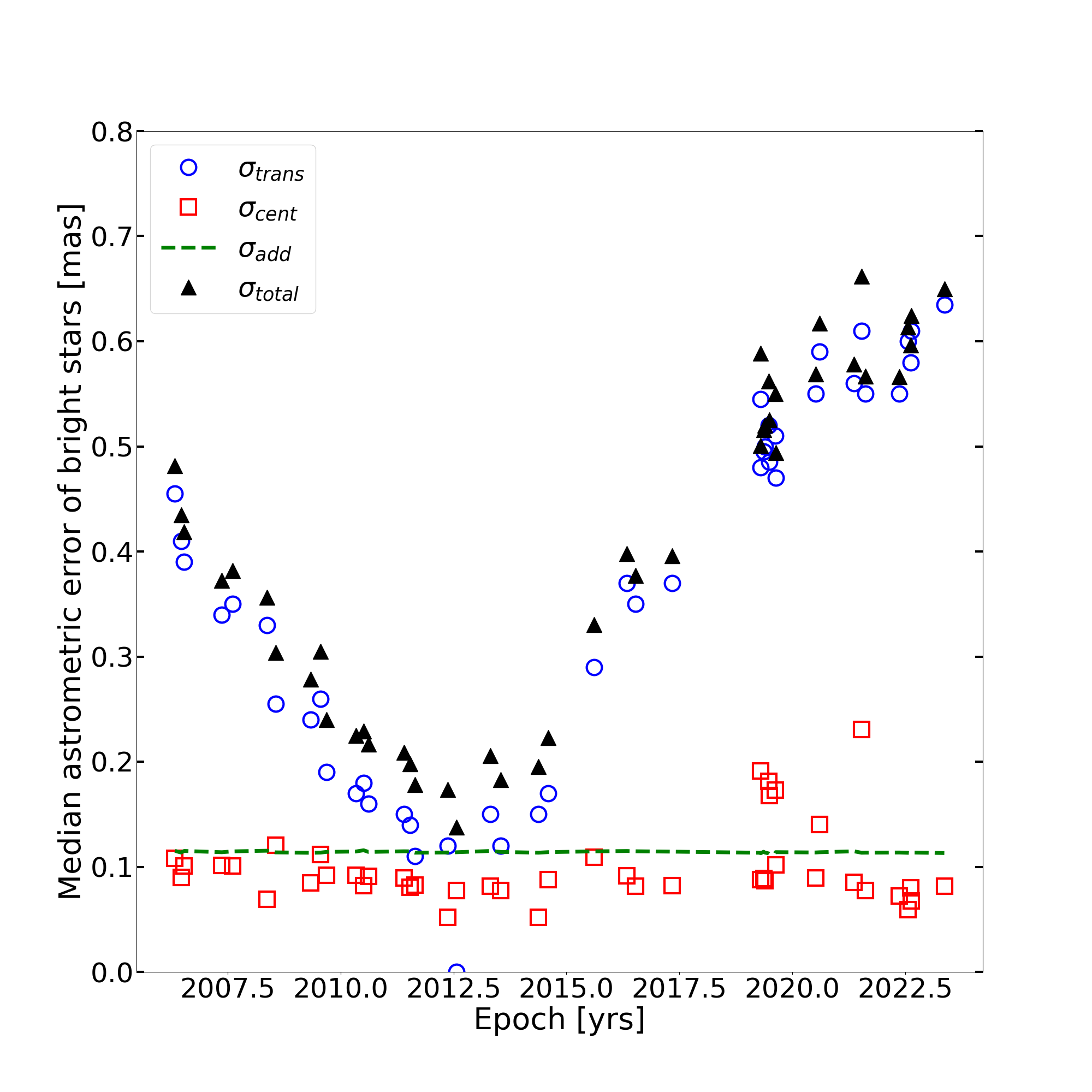}
    \caption{Median astrometric errors of bright stars in the Keck AO star lists. Transformation errors $\sigma_{\mathrm{trans}}$ are shown as blue circles, centroiding errors $\sigma_{\mathrm{cent}}$ are red squares, and additive errors from the \citet{Jia_2019} function are shown as a green dashed line. The total astrometric errorS $\sigma_{\mathrm{total}}$ with all these components quadrature summed together are shown as black triangles. Transformation errors originate from the transformation to Gaia-CRF3 with the 32 final reference stars. Bright stars here are Keck AO stars with \textit{K}$^{\prime}$ $<$ 15 mag and a radial offset from Sgr A*-radio of $<$ 4". }
    \label{fig:final_err_vs_epoch}
\end{figure}

The positional uncertainties in the Keck AO star lists have both statistical and systematic components \citep{Jia_2019}. The statistical component $\sigma_{\mathrm{cent}}$ reflects the centroiding uncertainty of the star's point spread function (PSF) fit across the three submaps for a given epoch. The systematic component $\sigma_{\mathrm{add}}$ is a magnitude-dependent additive error term derived by \citet{Jia_2019}. The underlying cause of this derived term is thought to be inaccuracies in the estimates of the PSF wings of neighboring sources during source extraction. The total positional uncertainty of a given star is therefore a quadrature-sum of the statistical and systematic components such that $\sigma_{\mathrm{pos}} = \sqrt{\sigma_{\mathrm{cent}}^{2} + \sigma_{\mathrm{add}}^{2}}$. The centroid uncertainties for bright stars ($\sim0.09$ mas) are typically smaller than the additive error ($\sim0.1$ mas). Figure \ref{fig:final_err_vs_epoch} shows these error components for different epochs of observations. 

The photometric uncertainties of the Keck AO observations also have two components \citep{Gautum_2019,Jia_2019}. First is the instrumental magnitude uncertainty $\sigma_{m}$. It is obtained from the variance of the photometry for a star across the three submaps in a given observation epoch \citep{Gautum_2019}. Then there is the zero-point correction uncertainty $\sigma_{ZP}$ from the absolute photometric calibration that converts the instrumental magnitudes to absolute magnitudes \citep{Gautum_2019}. The total photometric uncertainty for a given star in the Keck AO star lists in a single epoch is therefore $\sigma_{mag} = \sqrt{\sigma_{m}^{2} + \sigma_{ZP}^{2}}$.

\begin{longrotatetable}
\begin{deluxetable*}{@{\extracolsep{4pt}}llcccccc}
\label{tab:sgra_nir_tab}
\tabletypesize{\footnotesize}
\tablecolumns{10} 
\tablecaption{Sgr A* Near-Infrared Measurements}
\tablehead{
 \multicolumn{2}{c}{Date} & \colhead{K$^{\prime}$} & \colhead{R.A.} & \colhead{DEC} & \colhead{$\alpha^{*}$ - $\alpha^{*}_{\mathrm{Sgr A^{*}-radio}}$} & \colhead{$\delta$ - $\delta_{\mathrm{Sgr A^{*}-radio}}$\tablenotemark{a}} & \colhead{Sgr A*-IR Status}\\
 \cline{1-2}
\colhead{(U.T.)} & \colhead{(Decimal)} & \colhead{(mag)}  & \colhead{(deg)} & \colhead{(deg)} & \colhead{(mas)} & \colhead{(mas)} & \colhead{}
}
\startdata
2006 June 20-21 & 2006.470 & & & & & & Confused \\
2006 July 17 & 2006.541 & & & & & & Confused \\
2007 May 17 & 2007.374 & & & & & & Confused \\
2007 Aug 10-12 & 2007.610 & & & & & & Confused \\
2008 May 15 & 2008.370 & & & & & & Confused \\
2008 July 24 & 2008.562 & & & & & & Confused \\
2009 May 1-4  & 2009.336 & & & & & & Confused \\
2009 July 22-24 & 2009.561 & & & & & & Confused \\
2009 Sept 09 & 2009.689 & & & & & & Confused \\
2010 May 4-5 & 2010.340 & 15.850 $\pm$ 0.074  & 266.41681395 $\pm$ 0.00000021 & -29.00783053 $\pm$ 0.00000016 & 2.16 $\pm$ 0.50 & -1.28 $\pm$ 0.74 & Good detection \\
2010 July 06 & 2010.511 & 16.454 $\pm$ 0.069 & 266.41681410 $\pm$ 0.00000069 & -29.00782977 $\pm$ 0.00000040 & 3.2 $\pm$ 1.3 & 2.4 $\pm$ 2.4 & Good detection\\
2010 Aug 15 & 2010.620 & & & & & & Undetected \\ 
2011 May 27 & 2011.401 & 16.334 $\pm$ 0.057  & 266.41681339 $\pm$ 0.00000019 & -29.00783227 $\pm$ 0.00000019 & 3.72 $\pm$ 0.58 & -1.60 $\pm$ 0.66 & Good detection\\
2011 July 18 & 2011.543 & 17.127 $\pm$ 0.063  & 266.41681227 $\pm$ 0.00000025 & -29.00783246 $\pm$ 0.00000033 & 0.7 $\pm$ 1.0 & -1.48 $\pm$ 0.90 & Good detection\\
2011 Aug 23-24 & 2011.643 & 17.112 $\pm$ 0.032  & 266.41681327 $\pm$ 0.00000019 & -29.00783352 $\pm$ 0.00000042 & 4.1 $\pm$ 1.3 & -4.76 $\pm$ 0.72 & Good detection\\
2012 May 15-18 & 2012.372 & 17.974 $\pm$ 0.045 & 266.41681071 $\pm$ 0.00000087 & -29.0078349 $\pm$ 0.0000026 & -1.7 $\pm$ 8.2 & -5.8 $\pm$ 3.2 & Good detection\\
2012 July 24 & 2012.562 & 16.789 $\pm$ 0.058 & 266.41681138 $\pm$ 0.00000025 & -29.00783429 $\pm$ 0.00000058 & 1.1 $\pm$ 1.8 & -2.40 $\pm$ 0.94 & Good detection\\
2013 Apr 26-27 & 2013.319 & & & & & & Undetected \\ 
2013 July 20 & 2013.550 & 17.50 $\pm$ 0.32 & 266.4168085 $\pm$ 0.0000015 & -29.0078348 $\pm$ 0.0000021 & -5.0 $\pm$ 6.5 & 1.5 $\pm$ 5.4 & Good detection\\
2014 May 19 & 2014.380 & & & & & & Confused \\
2014 Aug 3-6 & 2014.595 & & & & & & Confused \\
2015 Aug 9-11 & 2015.607 & & & & & & Undetected \\ 
2016 May 3 & 2016.338 & & & & & & Confused \\
2016 July 13 & 2016.532 & & & & & & Confused \\
2017 May 4-5 & 2017.343 & & & & & & Confused \\
2019 Apr 19 & 2019.298 & & & & & & Undetected \\ 
2019 April 20 & 2019.300 & 15.501 $\pm$ 0.082 & 266.41680525 $\pm$ 0.00000023 & -29.00784419 $\pm$ 0.00000023 & 3.00 $\pm$ 0.71 & -0.40 $\pm$ 0.86 & Good detection\\
2019 May 13 & 2019.363 & 14.11 $\pm$ 0.11 & 266.41680485 $\pm$ 0.00000017 & -29.00784462 $\pm$ 0.00000016 & 1.95 $\pm$ 0.48 & -1.60 $\pm$ 0.65 & Good detection\\
2019 May 23 & 2019.391 & 15.95 $\pm$ 0.13 & 266.41680548 $\pm$ 0.00000034 & -29.00784467 $\pm$ 0.00000023 & 4.02 $\pm$ 0.70 & -1.6 $\pm$ 1.2 & Good detection\\
2019 June 25 & 2019.481 & & & & & & Undetected \\ 
2019 June 30 & 2019.495 & & & & & & Undetected \\ 
2019 Aug 14 & 2019.617 & 16.394 $\pm$ 0.080 & 266.41680443 $\pm$ 0.00000034 & -29.00784631 $\pm$ 0.00000059 & 1.4 $\pm$ 1.9 & -6.2 $\pm$ 1.3 & Good detection\\
2019 Aug 18-19 & 2019.630 & 16.74 $\pm$ 0.12 & 266.4168056 $\pm$ 0.0000013 & -29.00784519 $\pm$ 0.00000056 & 5.2 $\pm$ 1.7 & -2.2 $\pm$ 4.7 & Good detection\\
2020 July 07 & 2020.515 & & & & & & Undetected \\ 
2020 Aug 9 & 2020.606 & 15.857 $\pm$ 0.066 & 266.41680423 $\pm$ 0.00000058 & -29.00784634 $\pm$ 0.00000039 & 3.9 $\pm$ 1.2 & -0.8 $\pm$ 2.1 & Good detection\\
2021 May 13-14 & 2021.367 & & & & & & Confused \\
2021 July 13-14 & 2021.532 & & & & & & Confused \\
2021 Aug 12-14 & 2021.616 & & & & & & Confused \\ %
2022 May 14-15 & 2022.367 & & & & & & Confused \\
2022 July 19-22 & 2022.553 & & & & & & Confused \\
2022 Aug 14-16 & 2022.621 & & & & & & Confused \\
2022 Aug 19-20 & 2022.632  & & & & & & Confused \\
2023 May 12-13 & 2023.363 & & & & & & Confused \\
\enddata
\tablenotetext{a}{Positions noted as ``subtracted from Sgr A*-radio" are relative to Sgr A*-radio as measured by \citet{Xu_2022}.}
\end{deluxetable*}
\end{longrotatetable}

\subsection{HST-Gaia Catalog}
\label{subsec:hst_gaia_catalog}

We use reference sources from the HST-Gaia catalog produced by \citet{Hosek_2025} to convert our image pixel coordinates to sky coordinates. Out of the total 2876 reference stars in the HST-Gaia catalog, 235 fall within the FOV of our data (Figure \ref{fig:selected_ref_stars}). Of these stars, about two-thirds are brighter than \textit{K}$^{\prime}$ = 15 mag. The positional uncertainties of these stars, with both directions added in quadrature, are between 0.14 and 3.19 mas for bright stars (\textit{K}$^{\prime}$ $<$ 15 mag) and are between 0.43 and 5.19 mas for faint stars (\textit{K}$^{\prime}$ $>$ 15 mag), at the 5th and 95th percentiles, respectively. The proper motion uncertainties are between 0.03 and 0.61 mas yr$^{-1}$ for bright stars and between 0.09 and 1.07 mas yr$^{-1}$ for faint stars, at the 5th and 95th percentiles, respectively.

\section{Methods}
\label{sec:methods}

\subsection{Transforming Keck Adaptive Optics Astrometry into Gaia-CRF3}
\label{subsec:trans_keck_to_gaia}

To extract the Sgr A*-IR astrometry, we transform the pixel positions of the Keck AO star lists into Gaia-CRF3 sky coordinates. We follow the same procedure for the transformation as the one described at the beginning of Section 3 in \citet{Jia_2019}. But, the key differences between this work and \citet{Jia_2019} are our different source of reference stars and the different reference frame into which we are transforming the Keck AO observations. To start the transformation, we first select a reference epoch. The one chosen in this paper is 2012 July 24. This epoch has Keck AO astrometry with FWHM $\sim$ 58 mas and is near the middle of our Keck AO datasets in time. We then transform all other observation epochs into the reference epoch's coordinate system in pixel units by fitting a second-order bivariate polynomial transformation to the reference stars. Finally, we calculate the transformation from the reference-epoch pixel coordinate system to the reference frame in Gaia-CRF3. The reference epoch to reference frame transformation is applied to all observation epochs. 

In addition, for the Sgr A*-IR points, we correct beyond the global geometric distortion. For this, we use the central arcsecond local distortion correction method in the appendix of \citet{Jia_2019}. The method calculates the needed correction from correlations between the linear, accelerating, and orbital motion fits to stars within 1" of Sgr A*. The astrometric distortions that we correct for here result from different experimental setups for the observations that are of higher order than global geometric distortions as measured for NIRC2 by \citet{Yelda_2010} and \citet[][see also \citet{Jia_2019}]{Service_2016}. The lower-order global geometric distortion corrections from \citet{Yelda_2010} and \citet{Service_2016} are implemented in the data reduction (Section \ref{subsec:keck_ao_obs}).

The transformation uncertainty is determined using a half-sample bootstrap (without replacement) over the reference stars. The bootstrap has 100 iterations. For each iteration, we calculate the transformation using a random one-half of the reference stars. The transformation uncertainty for each star is the standard deviation of its positions in the 100 bootstrapped reference-epoch transformations. This transformation uncertainty $\sigma_{\mathrm{trans}}$ is added in quadrature with the  positional uncertainty $\sigma_{\mathrm{pos}}$ to obtain the total astrometric uncertainty $\sigma_{\mathrm{total}}$ such that $\sigma_{\mathrm{total}} = \sqrt{\sigma_{\mathrm{trans}}^2 + \sigma_{\mathrm{pos}}^2}$. Specifically for our Sgr A*-IR points, there is the additional term for the residual optical distortion $\sigma_{\mathrm{dist}}$ such that its total astrometric error is $\sigma_{\mathrm{total}} = \sqrt{\sigma_{\mathrm{trans}}^{2} + \sigma_{\mathrm{pos}}^{2} + \sigma_{\mathrm{dist}}^{2}}$. 

We refine our candidate reference star list (Section \ref{subsec:hst_gaia_catalog}) with the aim of selecting reference sources that minimize transformation uncertainties as well as positional and proper motion biases in the transformation. Positional and proper motion biases here refer to erroneous offsets in the pixel units to sky coordinates transformation. 

Following from \citet{Hosek_2025}, we define the proper motion bias as 

\begin{equation}
\label{equ:prop_motion_bias}
    \Delta \mu = \frac{\Sigma_{i = 0}^{N} w_{i} (\mu_{i,ref} - \mu_{i,transformed})}{\Sigma_{i = 0}^{N} w_{i}}
\end{equation}

\noindent where $\mu_{i,ref}$ is the known proper motion of a star from the HST-Gaia catalog and $\mu_{i,transformed}$ is the measured proper motion of that same star after transformation to Gaia-CRF3 sky coordinates. 

We define the positional bias as 

\begin{equation}
\label{equ:pos_bias}
    \Delta p = \frac{\Sigma_{i = 0}^{N} w_{i} (p_{i,ref} - p_{i,transformed})}{\Sigma_{i = 0}^{N} w_{i}}
\end{equation}

\noindent where $p_{i,ref}$ is the propagated position of a star to time 2012.562 (2012 July 24) based on its known proper motion in the HST-Gaia catalog and $p_{i,transformed}$ is the propagated position of that same star based on its measured proper motion after transformation to Gaia-CRF3 sky coordinates.

The term $w_{i}$ is the quadrature-summed weight where

\begin{equation}
    w_{i} = \frac{1}{\sqrt{\sigma_{i,ref}^{2} + \sigma_{i,transformed}^{2}}}
\end{equation}

\noindent where $\sigma_{i,ref}$ is the uncertainty of $\mu_{i,ref}$ or $p_{i,ref}$ for the proper motion or positional bias, respectively. Likewise, this nomenclature also maps to $\mu_{i,transformed}$ or $p_{i,transformed}$. The precision of the transformation biases are the weighted error on the mean of the proper motion or position differences used to calculate the biases.

Transformation errors are generally minimized by using many reference stars with small astrometric uncertainties. However, there is a trade-off between the number of reference sources and their astrometric uncertainties. There are many more faint stars with higher uncertainties than bright ones with low uncertainties. 

\begin{figure*}
    \centering
    \includegraphics[width=1\linewidth]{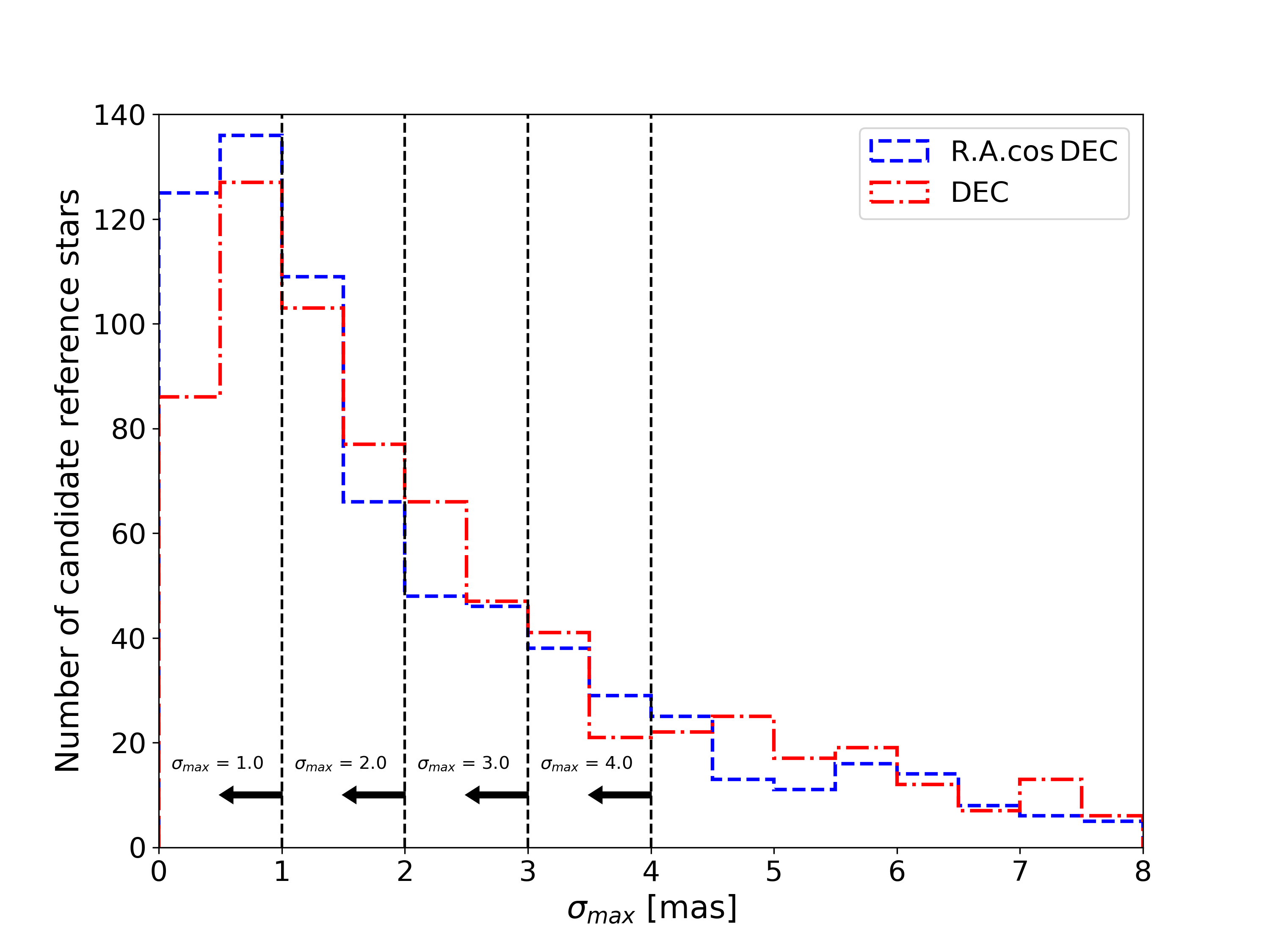}
    \caption{Maximum projected astrometric error for candidate reference stars from HST-Gaia catalog in R.A.$\cos{\mathrm{DEC}}$ (blue; dashed line) and DEC (red; dashed-dotted line). The dotted vertical lines denote the limits used to create the candidate reference star subsets.}
    \label{fig:candidate_ref_star_errors}
\end{figure*}

\begin{figure}
    \centering
    \includegraphics[width=1\linewidth]{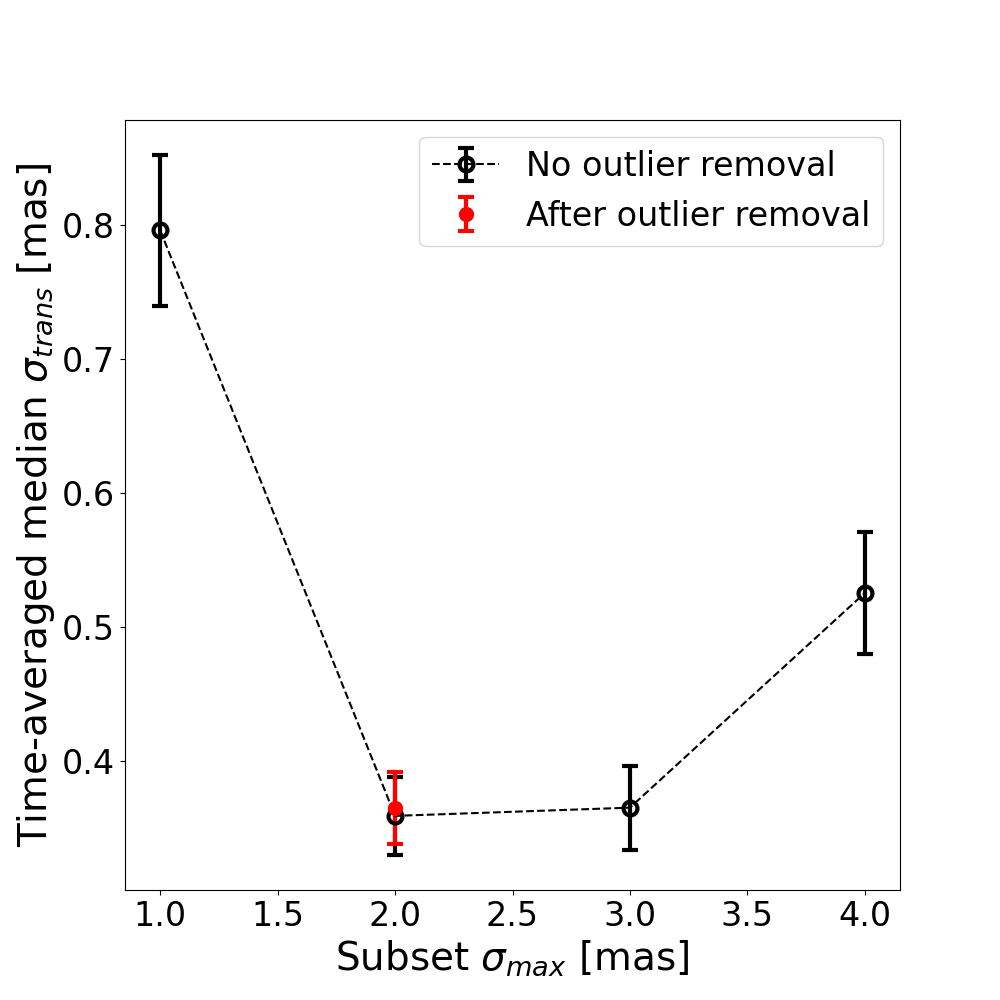}
    \caption{Time-averaged median transformation errors of Keck AO bright stars in transformations using the candidate star subsets. These averaged medians (unfilled black circles) are plotted as a function of the $\sigma_{\mathrm{max}}$ of the subset used in the transformation. Bright stars used here are stars with \textit{K} $<$ 15 mag and a radial offset from Sgr A*-radio of $<$ 4". The subset that minimized transformation errors had $\sigma_{\mathrm{max}}$ = 2.0 mas. Its time-averaged median $\sigma_{\mathrm{trans}}$ is shown in red as a filled circle after outlier reference stars were removed. Errorbars on the time-averaged median transformation errors are 1$\sigma$ of the distributions of the median transformation errors in each epoch.}
    \label{fig:trans_errs_per_subset}
\end{figure}

We test different subsets of candidate reference stars that are selected based on their projected positional uncertainties. For each reference star, we calculate the maximum projected positional error ($\sigma_{max}$) across the AO epochs using the HST-Gaia catalog proper motions. We then create nested samples of reference stars based on $\sigma_{max}$, with each sample containing all reference stars with $\sigma_{max}$ less than the limiting value of the subset. We explore $\sigma_{max}$ limit values of 1.0, 2.0, 3.0, and 4.0 mas. These thresholds compared to the overall $\sigma_{max}$ values of the candidate stars are shown in Figure \ref{fig:candidate_ref_star_errors}. We ultimately find that the sample with $\sigma_{max}$ $<$ 2.0 mas minimizes the transformation error (Figure \ref{fig:trans_errs_per_subset}). This gives us 43 reference stars that bring the time-averaged median transformation error of the bright stars (\textit{K}$^{\prime}$ $<$ 15 mag; r $<$ 4 arcsec) to 0.34 mas.

From this set of reference stars, we then reject outliers to minimize positional and proper motion bias. Outliers are defined by their positional and proper motion differences between the HST-Gaia catalog and posttransformation Keck AO star lists in Gaia-CRF3. We omit reference stars from the subset with proper motion differences of $\geq 5\sigma$ compared to their HST-Gaia catalog values and $t_{0}$ positions discrepant by $\geq 10\sigma$. Here, $\sigma$ is the quadratic sum of the astrometric errors from the transformed Keck AO star list and and HST-Gaia catalog. After the outlier stars were omitted, the transformation was redone a final time with the final set of 32 reference stars.

The final set of 32 reference stars is listed in Table \ref{tab:secondary_ref_stars} and shown in Figure \ref{fig:selected_ref_stars}. These reference stars have \textit{K}$^{\prime}$-band magnitudes between 10.3 and 14.9 mag. Their positional uncertainties range between 0.17 and 0.65 mas and their proper motion uncertainties range between 0.04 and 0.14 mas yr$^{-1}$, at the 5th and 95th percentiles, respectively. The resulting median total astrometric error $\sigma_{total}$ among the Keck AO bright stars (\textit{K}$<$15; not only reference stars) ranges between 0.14 and 0.66 mas, as shown in Figure \ref{fig:final_err_vs_epoch}, and they are dominated by the transformation errors. The median transformation errors range between 0.10 and 0.63 mas, which are 4 times larger than the median $\sigma_{\mathrm{cent}}$ and 3 times larger than the median $\sigma_{\mathrm{add}}$ on average. These transformation uncertainties are also significantly larger than those in \citet{Jia_2019} by $\sim5$ times on average. Further discussion of the Gaia-CRF3 transformation uncertainties is in Appendix \ref{appendix:astrometric_assessment}.

\begin{deluxetable*}{@{\extracolsep{4pt}}llccccccc}
\rotate
\label{tab:secondary_ref_stars}
\tabletypesize{\footnotesize}
\tablecolumns{10} 
\tablecaption{Astrometric Reference Stars}
\tablehead{
 \colhead{Name} & \colhead{K$^{\prime}$\tablenotemark{a}} & \colhead{t$_{0}$} & \colhead{R.A at t$_{0}$} & \colhead{DEC at t$_{0}$} & \colhead{$\alpha^{*}$ - $\alpha^{*}_{\mathrm{Sgr A^{*}-radio}}$} & \colhead{$\delta$ - $\delta_{\mathrm{Sgr A^{*}-radio}}$\tablenotemark{b}} &
 \colhead{$\mu_{\alpha^{*}}$ - $\mu_{\alpha^{*},\,\mathrm{Sgr A^{*}-radio}}$} & \colhead{$\mu_{\delta}$ - $\mu_{\delta,\,\mathrm{Sgr A^{*}-radio}}$} \\
\colhead{} & \colhead{(mag)}  & \colhead{(years)} & \colhead{(deg)} & \colhead{(deg)} & \colhead{(arcsec)} & \colhead{(arcsec)} &
\colhead{(mas yr$^{-1}$)} & \colhead{(mas yr$^{-1}$)} 
}
\startdata
irs33E & 10.271 $\pm$ 0.005 & 2018.325 & 266.417047 $\pm$ 0.000031 & -29.008718 $\pm$ 0.000025 & 0.76205 $\pm$ 0.00011 & -3.150750 $\pm$ 0.000090 & 6.699 $\pm$ 0.025 & -1.420 $\pm$ 0.024\\
S2-17 & 10.712 $\pm$ 0.005 & 2018.118 & 266.417256 $\pm$ 0.000031 & -29.008366 $\pm$ 0.000025 & 1.41706 $\pm$ 0.00011 & -1.883810 $\pm$ 0.000090 & 9.028 $\pm$ 0.025 & -0.278 $\pm$ 0.023\\
irs13E1 & 10.770 $\pm$ 0.005 & 2019.324 & 266.41585 $\pm$ 0.00012 & -29.008313 $\pm$ 0.000069 & -3.01226 $\pm$ 0.00044 & -1.68800 $\pm$ 0.00025 & -3.473 $\pm$ 0.11 & -2.637 $\pm$ 0.066\\
S3-22 & 11.124 $\pm$ 0.005 & 2018.774 & 266.416706 $\pm$ 0.000039 & -29.00874 $\pm$ 0.00016 & -0.31170 $\pm$ 0.00014 & -3.22758 $\pm$ 0.00057 & 3.131 $\pm$ 0.033 & -1.51 $\pm$ 0.11\\
irs33N & 11.279 $\pm$ 0.005 & 2018.324 & 266.416806 $\pm$ 0.000042 & -29.008480 $\pm$ 0.000053 & 0.00238 $\pm$ 0.00015 & -2.29475 $\pm$ 0.00019 & 3.514 $\pm$ 0.033 & -6.097 $\pm$ 0.042\\
S4-207 & 11.378 $\pm$ 0.005 & 2018.328 & 266.415522 $\pm$ 0.000028 & -29.00842 $\pm$ 0.00012 & -4.04174 $\pm$ 0.00010 & -2.08997 $\pm$ 0.00042 & 1.360 $\pm$ 0.026 & -1.625 $\pm$ 0.087\\
S1-23 & 11.706 $\pm$ 0.005 & 2017.898 & 266.416533 $\pm$ 0.000042 & -29.00827 $\pm$ 0.00010 & -0.85992 $\pm$ 0.00015 & -1.53123 $\pm$ 0.00036 & 4.482 $\pm$ 0.036 & -3.799 $\pm$ 0.076\\
S3-5 & 11.974 $\pm$ 0.005 & 2018.246 & 266.417752 $\pm$ 0.000083 & -29.008152 $\pm$ 0.000094 & 2.98058 $\pm$ 0.00030 & -1.11299 $\pm$ 0.00034 & 2.264 $\pm$ 0.074 & 4.750 $\pm$ 0.085\\
S4-4 & 12.033 $\pm$ 0.005 & 2018.168 & 266.417956 $\pm$ 0.000047 & -29.008535 $\pm$ 0.000056 & 3.61974 $\pm$ 0.00017 & -2.49311 $\pm$ 0.00020 & 1.644 $\pm$ 0.044 & -6.504 $\pm$ 0.051\\
S4-129 & 12.136 $\pm$ 0.005 & 2019.208 & 266.417987 $\pm$ 0.000053 & -29.00845 $\pm$ 0.00013 & 3.72301 $\pm$ 0.00019 & -2.19019 $\pm$ 0.00048 & 3.147 $\pm$ 0.046 & 3.500 $\pm$ 0.097\\
S2-32 & 12.319 $\pm$ 0.005 & 2017.835 & 266.417162 $\pm$ 0.000056 & -29.007066 $\pm$ 0.000047 & 1.12199 $\pm$ 0.00020 & 2.79133 $\pm$ 0.00017 & -0.367 $\pm$ 0.044 & 1.450 $\pm$ 0.038\\
S6-27 & 12.340  $\pm$ 0.005 & 2017.961 & 266.418099 $\pm$ 0.000058 & -29.009134 $\pm$ 0.000067 & 4.07137 $\pm$ 0.00021 & -4.65226 $\pm$ 0.00024 & -0.803 $\pm$ 0.052 & 3.326 $\pm$ 0.060\\
S3-26 & 12.384 $\pm$ 0.005 & 2017.162 & 266.416003 $\pm$ 0.000036 & -29.008414 $\pm$ 0.000036 & -2.52890 $\pm$ 0.00013 & -2.06412 $\pm$ 0.00013 & 5.916 $\pm$ 0.030 & 1.285 $\pm$ 0.029\\
S3-374 & 12.463  $\pm$ 0.005 & 2017.846 & 266.415930 $\pm$ 0.000042 & -29.008644 $\pm$ 0.000058 & -2.75715 $\pm$ 0.00015 & -2.88830 $\pm$ 0.00021 & -0.411 $\pm$ 0.035 & -4.628 $\pm$ 0.048\\
S4-2 & 12.704 $\pm$ 0.008 & 2018.692 & 266.41800 $\pm$ 0.00010 & -29.007390 $\pm$ 0.000072 & 3.77765 $\pm$ 0.00037 & 1.63068 $\pm$ 0.00026 & 1.477 $\pm$ 0.079 & -2.059 $\pm$ 0.059\\
S4-6 & 12.760 $\pm$ 0.005 & 2018.241 & 266.417854 $\pm$ 0.000056 & -29.008614 $\pm$ 0.000056 & 3.30049 $\pm$ 0.00020 & -2.77622 $\pm$ 0.00020 & 2.150 $\pm$ 0.047 & -3.080 $\pm$ 0.046\\
S2-74 & 13.253  $\pm$ 0.005 & 2017.725 & 266.416821 $\pm$ 0.000092 & -29.007068 $\pm$ 0.000094 & 0.04888 $\pm$ 0.00033 & 2.78472 $\pm$ 0.00034 & -8.775 $\pm$ 0.075 & 0.967 $\pm$ 0.076\\
S5-211 & 13.313 $\pm$ 0.005 & 2018.649 & 266.418218 $\pm$ 0.000064 & -29.008918 $\pm$ 0.000067 & 4.45009 $\pm$ 0.00023 & -3.86885 $\pm$ 0.00024 & -0.672 $\pm$ 0.051 & 4.188 $\pm$ 0.055\\
S3-149 & 13.353  $\pm$ 0.005 & 2019.062 & 266.415890 $\pm$ 0.000058 & -29.007354 $\pm$ 0.000072 & -2.87816 $\pm$ 0.00021 & 1.76206 $\pm$ 0.00026 & 3.038 $\pm$ 0.055 & 6.729 $\pm$ 0.069\\
S3-7 & 13.623  $\pm$ 0.005 & 2018.068 & 266.417414 $\pm$ 0.000097 & -29.008572 $\pm$ 0.000094 & 1.91442 $\pm$ 0.00035 & -2.62623 $\pm$ 0.00034 & -0.831 $\pm$ 0.086 & -1.189 $\pm$ 0.085\\
S3-370 & 13.633 $\pm$ 0.005 & 2018.291 & 266.41672 $\pm$ 0.00012 & -29.00892 $\pm$ 0.00012 & -0.28387 $\pm$ 0.00043 & -3.88495 $\pm$ 0.00043 & 0.61 $\pm$ 0.10 & 4.40 $\pm$ 0.10\\
S3-198 & 13.726 $\pm$ 0.005 & 2018.880 & 266.417011 $\pm$ 0.000056 & -29.00689 $\pm$ 0.00012 & 0.65070 $\pm$ 0.00020 & 3.43013 $\pm$ 0.00042 & -0.335 $\pm$ 0.047 & -1.653 $\pm$ 0.094\\
S4-287 & 13.734 $\pm$ 0.005 & 2019.614 & 266.416856 $\pm$ 0.000033 & -29.009167 $\pm$ 0.000086 & 0.16287 $\pm$ 0.00012 & -4.76121 $\pm$ 0.00031 & 2.957 $\pm$ 0.039 & 1.146 $\pm$ 0.085\\
S4-59 & 13.828 $\pm$ 0.005 & 2018.717 & 266.415808 $\pm$ 0.000031 & -29.007110 $\pm$ 0.000053 & -3.13782 $\pm$ 0.00011 & 2.64035 $\pm$ 0.00019 & 1.841 $\pm$ 0.028 & -0.058 $\pm$ 0.045\\
S2-2 & 14.046 $\pm$ 0.005 & 2018.670 & 266.416646 $\pm$ 0.000061 & -29.00725 $\pm$ 0.00010 & -0.49910 $\pm$ 0.00022 & 2.12596 $\pm$ 0.00037 & 2.454 $\pm$ 0.049 & 4.325 $\pm$ 0.085\\
S3-207 & 14.056 $\pm$ 0.005 & 2018.581 & 266.417303 $\pm$ 0.000064 & -29.006947 $\pm$ 0.000061 & 1.56926 $\pm$ 0.00023 & 3.22616 $\pm$ 0.00022 & 2.514 $\pm$ 0.057 & 2.473 $\pm$ 0.056\\
S4-221 & 14.245 $\pm$ 0.005 & 2017.963 & 266.41768 $\pm$ 0.00016 & -29.008899 $\pm$ 0.000094 & 2.73993 $\pm$ 0.00056 & -3.80478 $\pm$ 0.00034 & 6.78 $\pm$ 0.11 & -5.021 $\pm$ 0.073\\
S5-43 & 14.311 $\pm$ 0.005 & 2018.929 & 266.41739 $\pm$ 0.00011 & -29.00918 $\pm$ 0.00011 & 1.82827 $\pm$ 0.00040 & -4.81653 $\pm$ 0.00038 & -1.804 $\pm$ 0.090 & -2.272 $\pm$ 0.087\\
S5-198 & 14.642 $\pm$ 0.005 & 2017.654 & 266.416978 $\pm$ 0.000089 & -29.00945 $\pm$ 0.00016 & 0.54262 $\pm$ 0.00032 & -5.79967 $\pm$ 0.00056 & 3.897 $\pm$ 0.072 & 2.49 $\pm$ 0.12\\
S5-83 & 14.702 $\pm$ 0.006 & 2016.472 & 266.418453 $\pm$ 0.000072 & -29.00811 $\pm$ 0.00016 & 5.18132 $\pm$ 0.00026 & -0.98484 $\pm$ 0.00058 & -3.396 $\pm$ 0.060 & -6.04 $\pm$ 0.13\\
S2-69 & 14.767 $\pm$ 0.005 & 2018.165 & 266.41652 $\pm$ 0.00013 & -29.00712 $\pm$ 0.00013 & -0.89149 $\pm$ 0.00046 & 2.59647 $\pm$ 0.00046 & -1.811 $\pm$ 0.092 & 5.226 $\pm$ 0.092\\
S4-98 & 14.847 $\pm$ 0.005 & 2019.855 & 266.415486 $\pm$ 0.000075 & -29.008082 $\pm$ 0.000069 & -4.14849 $\pm$ 0.00027 & -0.85296 $\pm$ 0.00025 & -2.883 $\pm$ 0.081 & 5.522 $\pm$ 0.078\\
\enddata
\tablenotetext{a}{\textit{K}-band magnitudes are weighted averages across the Keck AO observations in which they are detected but not confused with a nearby star. The corresponding uncertainties are weighted errors on the mean.}
\tablenotetext{b}{Positions and proper motions noted as ``subtracted from Sgr A*-radio" are relative to Sgr A*-radio as measured by \citet{Xu_2022}.}
\end{deluxetable*}

\subsection{Astrometry and Motion Fit of the Infrared Counterpart to Sgr A*}
\label{subsec:sgra_motion}

We fit for the motion of Sgr A*-IR using both a first- and a second-order polynomial model on its observed positions in Gaia-CRF3. Out of our 44 observation epochs, our data reduction pipeline automatically made 25 detections of Sgr A*-IR. However, we used the results from star-planting simulations performed by \citet{Weldon_2023} and \citet{Paugnat_2024} to determine in which epochs the position of Sgr A*-IR is confused or heavily biased by a nearby star. These simulations found that the Sgr A*-IR astrometry would be compromised if a nearby star comes within $\sim70$ mas \citep[e.g. Table 9 from][]{Weldon_2023}. Thus, we limited our sample to measurements where no known star comes within that limit. This resulted in the removal of 11 supposed detections. After the confusion removal, we have 14 usable Sgr A*-IR positional measurements (Table \ref{tab:sgra_nir_tab}). \footnote{Of note, source extraction in this work is performed in the same way as \citet{Do_2019} and not in an enhanced mode that used a priori knowledge of the location of Sgr A* as in \citet{Weldon_2023}. This resulted in six epochs of nondetections of Sgr A*-IR in this work compared to \citet{Weldon_2023}.} Our Sgr A*-IR detections are average \textit{K}$^{\prime}$-band magnitudes, encompassing states of flaring and nonflaring, over the course of each observation epoch. 

We fit the Sgr A*-IR astrometry using linear and accelerating kinematic models. These models use Gaussian processes to simultaneously fit the motion of the source and systematic correlations in the astrometry \citep{Hosek_2025}. Gaussian processes are a flexible statistical tool to model a range of physical processes either parametrically or nonparametrically \citep{Rasmussen_2006}. For this application, we use the implementation in \citet{Hosek_2025}, which defines kinematic models with a polynomial kernel combined with various systematic uncertainty kernels. We fit the Sgr A*-IR astrometry with a total of eight of these kinematic models. These models are the first- and second-order polynomial models alone (no systematic model) as well as each polynomial paired with one of three systematic uncertainty models, which are a squared-exponential kernel (for time-dependent systematics), a confusion kernel (for spatially dependent systematics), and a constant additive error. 

The best-fit kinematic model is determined using its expected log predictive density \citep[ELPD;][]{Gelman_2013, Vehtari2017}. An ELPD serves as a measure of residuals between a model's predicted positions and the inputted observed position values. It also takes into account the uncertainty on the predicted positions. This means the ELPD can be used as a model selection criteria \citep{Gelman_2013}. For a multicomponent kinematic model to be preferred over a single-component kinematic model, we require that the multicomponent model be preferred by a probability threshold of $99.7\%$ ($3\sigma$, $\Delta \mathrm{ELPD} \sim 6$). The calculation of these probabilities using the ELPD are described in Appendix \ref{appendix:elpds}.

\section{Results}
\label{sec:results}

\subsection{Systematic Biases in Transformation to Gaia-CRF3}
\label{subsec:stat_consist}

\begin{figure*}
    \plottwo{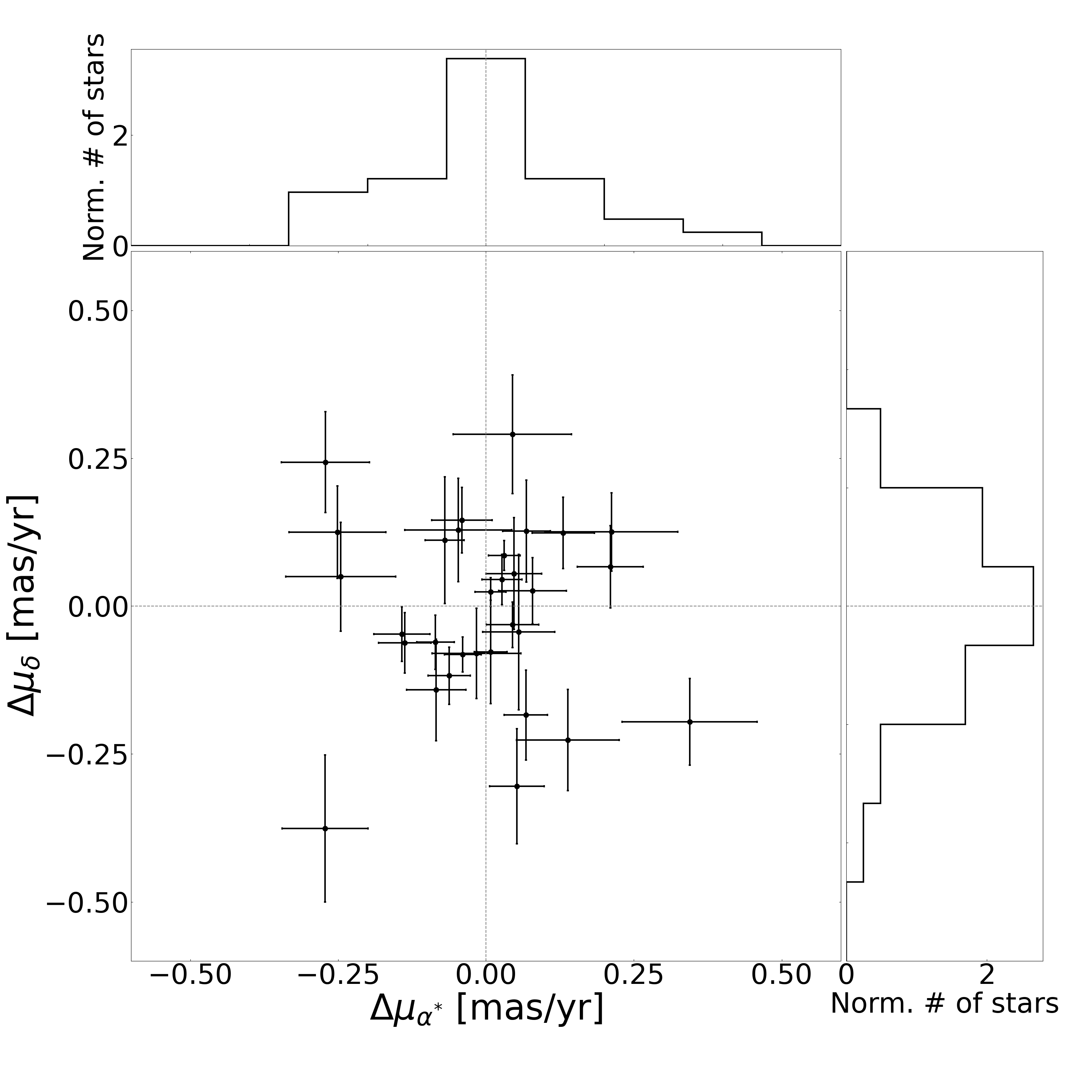}{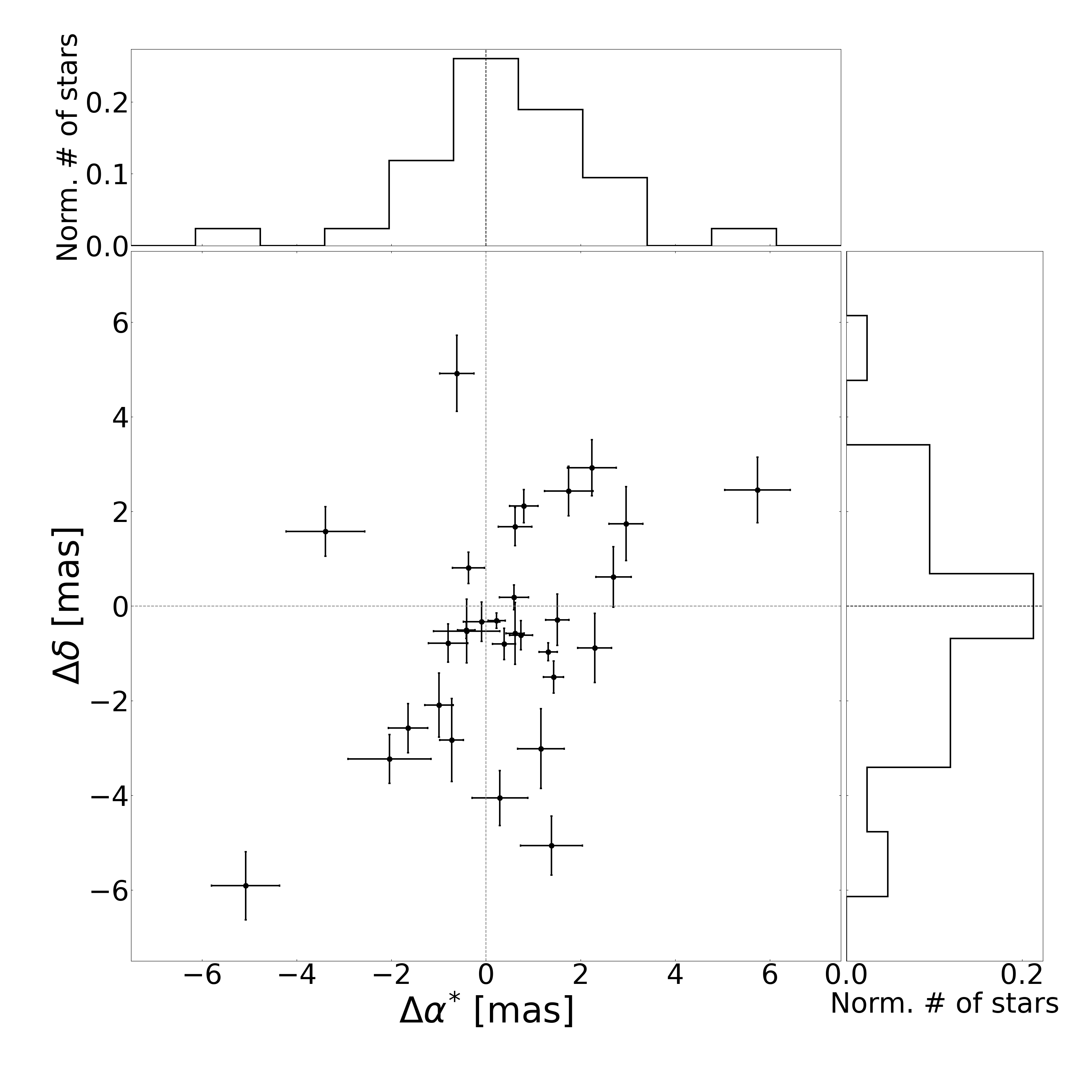}
    \caption{Differences of the final reference stars between the HST-Gaia catalog (already in Gaia-CRF3) and the Keck AO star lists after the transformations are applied. The left plot shows the differences in proper motion and right plot shows the differences in positional values at a common epoch (2012.562). The weighted mean proper motion biases are $\Delta\mu_{\alpha^{*}} = -0.0067 \pm 0.0079$ mas yr$^{-1}$ and $\Delta\mu_{\delta} = 0.003 \pm 0.010$ mas yr$^{-1}$. The weighted mean position biases are $\Delta\alpha^{*} = 0.538 \pm 0.055$ mas and $\Delta\delta = -0.390 \pm 0.066$ mas.}
    \label{fig:bias_plots}
\end{figure*}

We use the proper motion bias and position bias of the reference stars to characterize the quality of the transformation into Gaia-CRF3. The transformation uncertainties ($\sigma_{trans}$) represent the statistical uncertainties of each epoch, but the systematic uncertainties of the entire transformation are quantified by the biases. If the transformation of the Keck AO star lists into Gaia-CRF3 were perfect, we would expect the biases to be consistent with zero within the uncertainty. The biases represent the minimum systematic astrometric uncertainty that can be obtained in the transformed Keck AO star lists, including for Sgr A*-IR.

The differences in proper motion and position between the HST-Gaia catalog and the transformed Keck AO star lists are shown in Figure \ref{fig:bias_plots}. We find the proper motion biases for the reference stars are $\Delta\mu_{\alpha^{*}} = -0.0067 \pm 0.0079$ mas yr$^{-1}$ and $\Delta\mu_{\delta} = 0.003 \pm 0.010$ mas yr$^{-1}$. These indicate that there is no evidence for significant drift between the transformed Keck AO observations and the HST-Gaia catalog to a total proper motion precision of $\sqrt{\sigma^{2}_{\Delta\mu_{\alpha^{*}}} + \sigma^{2}_{\Delta\mu_{\delta}}} \sim 0.01$ mas yr$^{-1}$.

For position, we find that the positional biases are $\Delta\alpha^{*} = 0.538 \pm 0.055$ mas and $\Delta\delta = -0.390 \pm 0.066$ mas. Unlike the proper motions, we do find a bulk offset in the bias greater than its precision, making it inconsistent with zero. The total systematic bias in position is $\sqrt{(\Delta\alpha^{*})^{2}+ (\Delta\delta)^{2}} \sim 0.7$ mas. We assume that this is about the minimum systematic error that can be achieved by the astrometric transformation. We account for it when assessing the Sgr A*-IR kinematic fit.

\subsection{Kinematic Fit to the Infrared Counterpart to Sgr A*}
\label{subsec:sgra_ir_motion_fit}

\begin{deluxetable}{@{\extracolsep{4pt}}llc}
\label{tab:kinematic_models}
\tabletypesize{\footnotesize}
\tablecolumns{10} 
\tablecaption{Kinematic Models Tested}
\tablehead{
 \colhead{Polynomial} & \colhead{Systematic Error Model} & \colhead{Summed ELPD}\\
}
\startdata 
First order & None &  -71.8\\
            & Squared exponential & -60.2\\
            & Confusion & -60.1\\
            & Constant additive & -61.0\\
Second order & None & -71.5\\
             & Squared exponential  &  -60.2\\
             & Confusion  & -60.1\\
             & Constant additive & -60.8\\
\enddata
\end{deluxetable}

\begin{deluxetable*}{@{\extracolsep{4pt}}llcccc}
\label{tab:kinematic_fit_best}
\tabletypesize{\footnotesize}
\tablecolumns{10} 
\tablecaption{Parameters of Best Kinematic Fit}
\tablehead{
 \colhead{Parameter} & \colhead{Units} & \colhead{Value} & \colhead{Kinematic model uncertainty} & \colhead{Transformation uncertainties\tablenotemark{a}} & \colhead{Total uncertainty}\\
}
\startdata 
$\alpha^{*}(t = 2016.0)$ & deg & 266.416808480 & 0.00000025 & 0.00000015 & 0.00000029\\
$\delta(t = 2016.0)$ & deg &-29.007839465 & 0.00000048 & 0.00000011 & 0.00000027\\
$\mu_{\alpha^{*}}$ & mas yr$^{-1}$ &-3.093  & 0.082 & 0.022 & 0.085 \\
$\mu_{\delta}$ & mas yr$^{-1}$ & -5.62 & 0.13 &  0.018 & 0.13\\
\enddata

\tablenotetext{a}{Transformation uncertainties include the biases of transforming the HST-Gaia catalog into Gaia-CRF3 and of transforming the Keck AO star lists into the same reference frame as the HST-Gaia catalog.}
\end{deluxetable*}

We find that the preferred kinematic model for Sgr A*-IR astrometry is a first-order polynomial model paired with a constant systematic additive error. Table \ref{tab:kinematic_models} presents the eight kinematic model fits we performed on the Sgr A*-IR astrometry with corresponding ELPDs. Overall, we find that multicomponent models (polynomial model with systematic model) are preferred over single-component models (no systematic model). With our required preference probability threshold of $99.7\%$ ($3\sigma$), we cannot say that any one systematic model is preferred over the others. However, we take constant additive error as the preferred systematic uncertainty model because it is the simplest model for the systematic uncertainty. Selecting either the confusion or squared-exponential systematic error models instead of the constant additive model cause $\ll 0.1\%$ change in position and $<2\%$ change in proper motion. These are statistically insignificant changes to the position and proper motion fit.

\begin{figure}
    \gridline{\fig{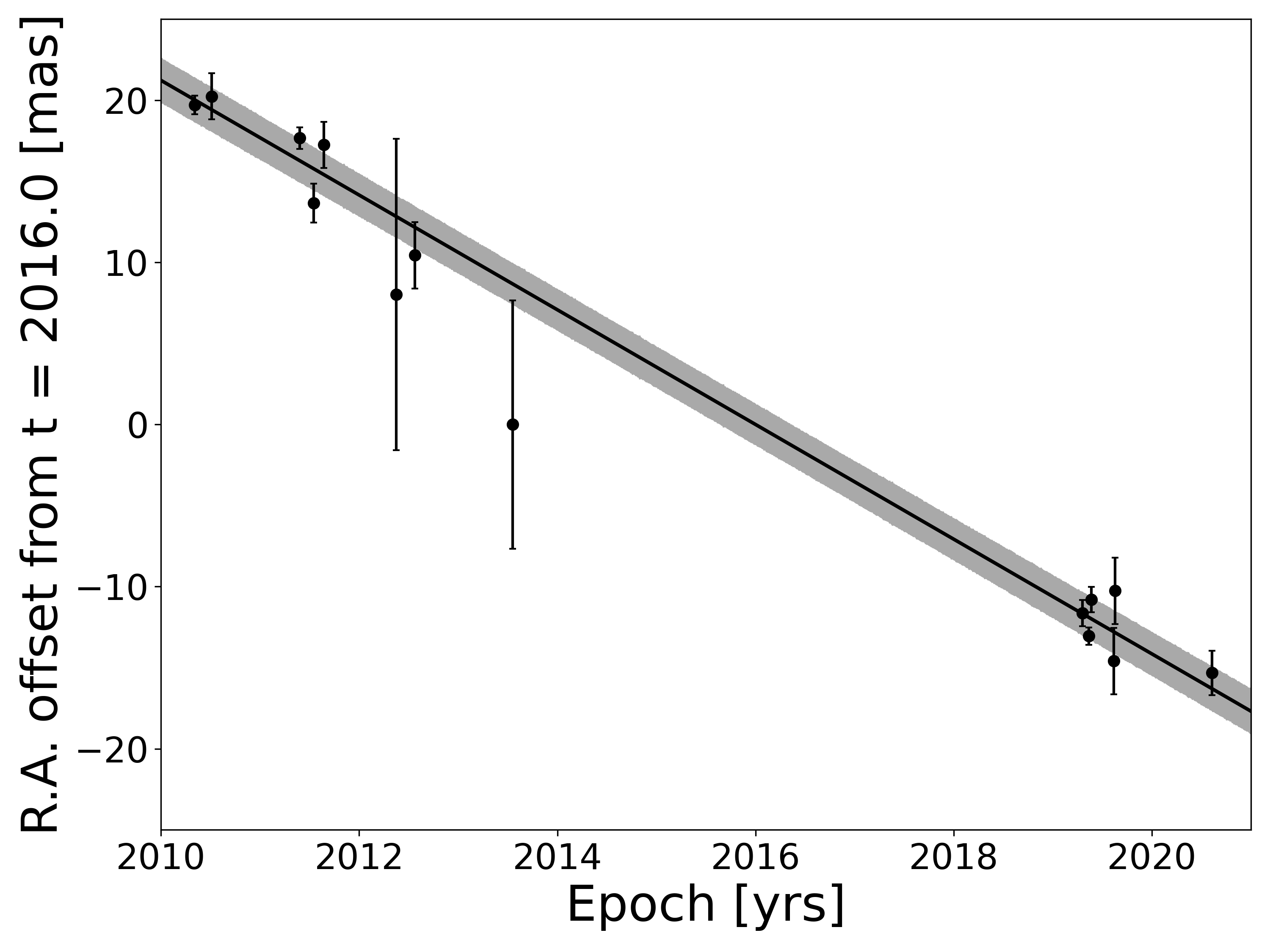}{0.45\textwidth}{(a)}}
    \gridline{\fig{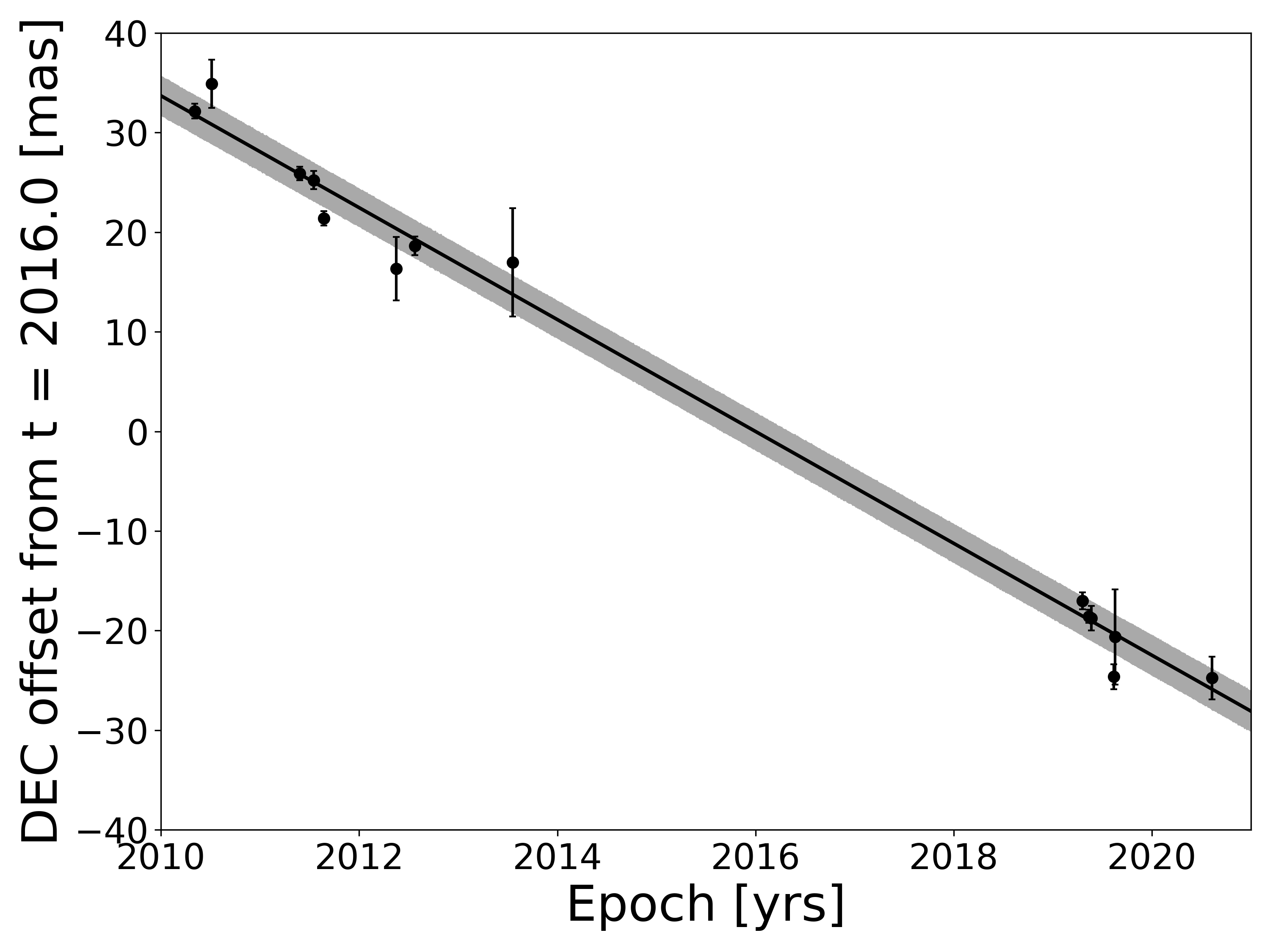}{0.45\textwidth}{(b)}}
    \caption{Kinematic motion fit of Sgr A*-IR, plotted as position offset from \textit{t} = 2016.0 as a function of time. Observed Sgr A*-IR positions are plotted as black circles with error bars and 1$\sigma$ error envelopes of the kinematic models are plotted in gray, with R.A. offset in (a) and DEC offset in (b). The position of Sgr A*-IR at \textit{t} = 2016.0 in Gaia-CRF3 is $\alpha$ = 266.41680848 $\pm$ 0.00000029 deg and $\delta$ = -29.00783947 $\pm$ 0.00000050 deg.}
    \label{fig:sgra_vs_time}
\end{figure}

\begin{figure*}
    \centering
    \includegraphics[width=1.0\linewidth]{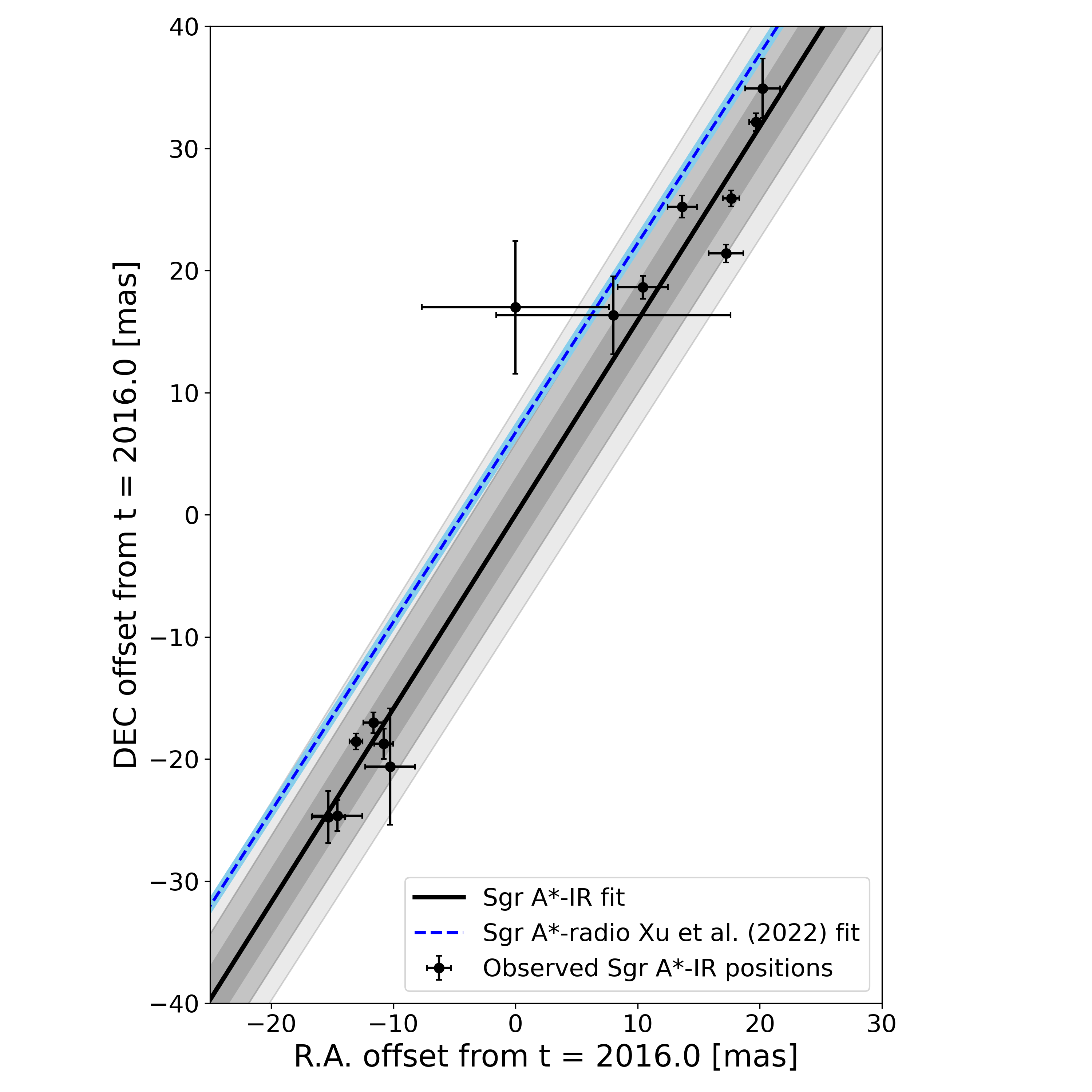}
    \caption{Kinematic motion fits of Sgr A*-IR and Sgr A*-radio plotted in R.A. and DEC as offset from the Sgr A*-IR fitted position at t = 2016.0. The kinematic motion fit to Sgr A*-IR from Table \ref{tab:kinematic_fit_best} is plotted as a solid black line with 1$\sigma$, $2\sigma$, and $3\sigma$ lighter gray error envelopes. Observed Sgr A*-IR positions described in this work are plotted as black circles with accompanying error bars. The motion fit to Sgr A*-radio from \citet{Xu_2022} is plotted as a dashed blue line with a 1$\sigma$ light blue error envelope. }
    \label{fig:sgra_2D}
\end{figure*}

We present the first-order preferred kinematic model for Sgr A*-IR in Table \ref{tab:kinematic_fit_best} and Figures \ref{fig:sgra_vs_time} and \ref{fig:sgra_2D}. The kinematic model is fit independently in R.A. and DEC with the $t_{0}$ of each fit at the weighted observation time in the respective directions. For clarity, all plots of the motion fits and given motion fit coefficients are reported at $t = 2016.0$. The polynomial component of the kinematic fit yields uncertainties in the position at t = 2016.0 for R.A.$\cos{\mathrm{DEC}}$ and DEC that are ($\sigma_{poly,\alpha^{*}}$, $\sigma_{poly,\delta}$) = (0.36, 0.56) mas and for the proper motion are ($\sigma_{\mu_{poly, \alpha^{*}}}$, $\sigma_{\mu_{poly,\delta}}$) = (0.082, 0.13) mas yr$^{-1}$. The systematic uncertainty component of the fit gives a constant additive error in each direction, and they are $\sigma_{sys,\alpha^{*}} = 0.82$ mas in R.A.$\cos{\mathrm{DEC}}$ and $\sigma_{sys,\delta} = 1.65$ mas in DEC. The systematic uncertainty is higher in the DEC-direction because, while the astrometry in both directions have similar scatter, the R.A.$\cos{\mathrm{DEC}}$ direction has enough uncertainty in the total astrometric positions $\sigma_{total}$ to compensate for it.

The total uncertainty of the kinematic coefficients for Sgr A*-IR in Gaia-CRF3 has components in addition to the model fit (Table \ref{tab:kinematic_fit_best}). First, there are the systematic uncertainties of transforming the Keck AO star lists into Gaia-CRF3. These are represented by the transformation biases, which we defined with equations \ref{equ:prop_motion_bias} and \ref{equ:pos_bias}. Because the position bias value was found to be larger than its precision, we take the entire position bias as the position systematic uncertainty. However, for the proper motion, we take the proper motion bias precision as the systematic uncertainty. This is because its offset is smaller than its precision in proper motion but not in position. We add these systematic uncertainties in quadrature with the model fit uncertainties.  In addition, we must include the systematic uncertainties on the transformation of the HST star lists to Gaia-CRF3 performed in \citet{Hosek_2025} to create the HST-Gaia catalog. These were calculated as transformation biases in the same manner as ours using equations \ref{equ:prop_motion_bias} and \ref{equ:pos_bias}. The proper motion biases of the HST-Gaia catalog in R.A.$\cos{\mathrm{DEC}}$ ($\sigma_{\Delta\mu_{\alpha^{*},HST}}$) and DEC ($\sigma_{\Delta\mu_{\delta},HST}$) are ($\sigma_{\Delta\mu_{\alpha^{*},HST}}$, $\sigma_{\Delta\mu_{\delta},HST}$) = (0.020, 0.015) mas yr$^{-1}$. The position biases in R.A.$\cos{\mathrm{DEC}}$ ($\sigma_{\Delta\alpha^{*},HST}$) and DEC ($\sigma_{\Delta\delta,HST}$) are ($\sigma_{\Delta\alpha^{*},HST}$, $\sigma_{\Delta\delta,HST}$) = (0.032, 0.030) mas. With all these components added in quadrature, our fitted proper motion for Sgr A*-IR is $\mu_{\alpha^{*}} = -3.093 \pm 0.085$ mas yr$^{-1}$ and $\mu_{\delta} = -5.62 \pm 0.13$ mas yr$^{-1}$. The $t = 2016.0$ position is ($\alpha(t)$, $\delta(t)$) = (266.41680848 $\pm$ 0.00000029, -29.00783947 $\pm$ 0.00000050) deg.  These positional uncertainties equate to 1.05 and 1.79 mas, respectively. We see that the total uncertainty in proper motion and position is dominated by our kinematic model fit uncertainties (Table \ref{tab:kinematic_fit_best}).

\section{Discussion}
\label{sec:discussion}

\subsection{Kinematics of the Infrared Counterpart to Sgr A* Compared to the Kinematics of the Radio Counterpart to Sgr A*}
\label{subsec: sgrair_vs_sgraradio}

\begin{figure*}
    \centering
    \includegraphics[width=1\linewidth]{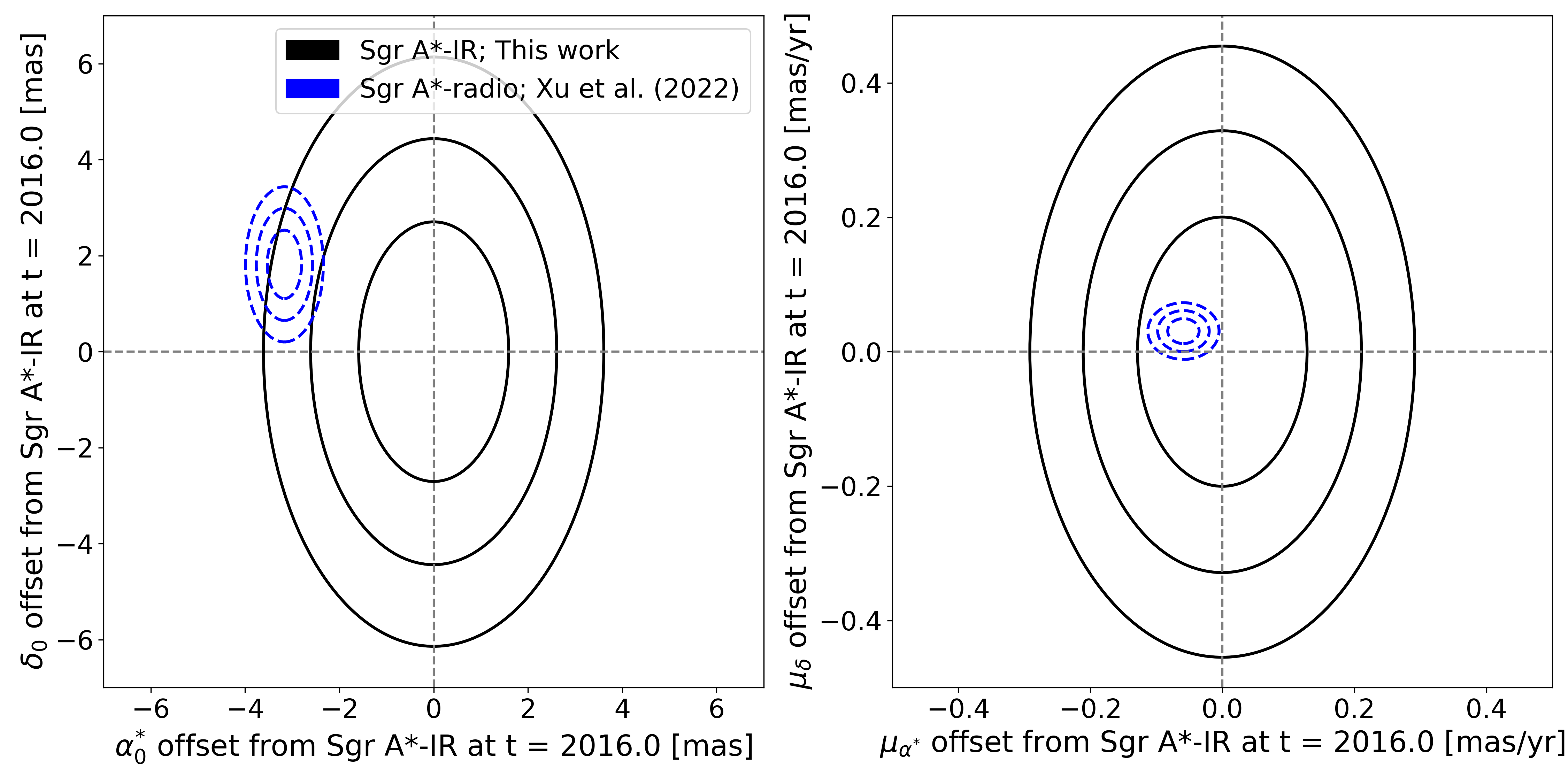}
    \caption{Contours of kinematic motion fits of Sgr A*-IR and Sgr A*-radio plotted in R.A.$\cos\mathrm{DEC}$ and DEC as offset from the Sgr A*-IR fitted position at \textit{t} = 2016.0. The fit to Sgr A*-radio as determined by \citet{Xu_2022} is plotted in blue as dashed lines and the preferred fit for Sgr A*-IR in this work is plotted in black as solid lines.}
    \label{fig:sgra_contours}
\end{figure*}

We find that our fit for the proper motion of Sgr A*-IR is consistent with the proper motion of Sgr A*-radio as measured by \citet[][Figures \ref{fig:sgra_2D} and \ref{fig:sgra_contours}]{Xu_2022}. \footnote{The proper motion for Sgr A*-radio reported by \citet{Xu_2022} was $\mu_{\alpha^{*}} = -3.152 \pm 0.011$ mas yr$^{-1}$ and $\mu_{\delta} = -5.586 \pm 0.006$ mas yr$^{-1}$.} The proper motion differences between Sgr A*-IR and Sgr A*-radio are, in R.A.$\cos{\mathrm{DEC}}$, $(\mu_{\alpha^{*},\,IR} - \mu_{\alpha^{*},\,radio}) = 0.059 \pm 0.082$ mas yr$^{-1}$ and, in DEC, $(\mu_{\delta,\,IR} - \mu_{\delta,\,radio}) = -0.03 \pm 0.13$ mas yr$^{-1}$. The uncertainties on these differences are the quadrature sum of the Sgr A*-IR total proper motion uncertainty (Section \ref{subsec:sgra_ir_motion_fit}), and the uncertainty between Sgr A*-radio and Gaia-CRF3. The uncertainty between Sgr A*-radio and Gaia-CRF3 is a quadrature sum of the uncertainty between Gaia-CRF3 and ICRF3 \citep[0.007 mas in position and $\sim 0.01$ mas yr$^{-1}$ in proper motion;][]{Hosek_2025} and the uncertainty of Sgr A*-radio in ICRF3 \citep[$\sim 0.5$ mas in position and $\sim 0.01$ mas yr$^{-1}$ in proper motion;][]{Xu_2022}. Overall, our uncertainties on the proper motion of Sgr A*-IR are about an order of magnitude larger than those on Sgr A*-radio. We believe this is predominantly due to the large amount of scatter in our observed Sgr A*-IR astrometry ($\sim2-3$ mas), which necessitates the need for the constant additive error kernel to appropriately increase the size of the fitted model's astrometric uncertainty. 

Our fit is also consistent with the Sgr A*-IR motion having zero acceleration, as does the Sgr A*-radio measurement by \citet{Reid_2020}. We compare our acceleration constraints to \citet{Reid_2020} because \citet{Xu_2022} did not make a constraint on the acceleration of Sgr A*-radio. Using our second-order polynomial fit with a constant additive error model, we can place a 2$\sigma$ upper limit on the acceleration on the sky of Sgr A*-IR of 0.061 mas yr$^{-2}$. \citet{Reid_2020} put a $2\sigma$ upper limit on the acceleration of Sgr A*-radio of $0.008$ mas yr$^{-2}$, which is an order of magnitude lower than our constraint. This is expected due to the high amount of scatter in the Sgr A*-IR points, likely caused by confusion with unresolved sources, which requires a large additional systematic uncertainty to compensate for it. Observations of Sgr A*-IR would need to be made with finer angular resolution to resolve the scattering.

In terms of position, we find that our derived position for Sgr A*-IR (at $t_0$ = 2016.0) is consistent with the Sgr A*-radio position within the uncertainties. Our preferred fit to Sgr A*-IR in Gaia-CRF3 has a systematic position offset of 3.31 $\pm$ 2.13 mas (at $t$ = 2016.0) from Sgr A*-radio when compared with the \citet{Xu_2022} measurements. However, this offset is not statistically significant at a difference of 1.6$\sigma$. The reason for this offset is believed to be systematics with aligning the Keck AO observations to the Gaia-CRF3 reference frame, not a physical mechanism.

\subsection{Constraints on Intermediate-mass Black Hole Parameters from the Acceleration Limits of the Infrared Counterpart to Sgr A*}
\label{subsec:imbh_limits}

We place constraints on the possible parameters for a hypothetical IMBH companion to the SMBH associated with Sgr A*. To make the constraints, we combine our Sgr A*-IR acceleration constraint from Section \ref{subsec:sgra_ir_motion_fit} with the average total astrometric error from the kinematic model fit, which is, across the time range of Sgr A*-IR detections, $\delta\theta \sim 2.07$ mas (for both the first- and second-order polynomial fits). These values allow us to create two bounds on the possible IMBH parameters in the mass versus orbital distance parameter space based on the limitations of our astrometric and kinematic measurements. They are adapted from \citet{Hansen_2003}.

For the first bound, if we assume a circular orbit for the IMBH companion around the barycenter between it and the SMBH, we can use the acceleration upper limit to exclude areas of mass versus orbital distance parameter space. In this regime, all allowed masses $M_{\mathrm{IMBH}}$ and orbital semi-major axes $r$ for the IMBH must have circular accelerations lower than this value. The allowed parameter space here is bounded by

\begin{equation}
    M_{\mathrm{IMBH}} < \frac{a_{\mathrm{sky}}R_{\mathrm{GC}}}{G}\,r^{2}
\end{equation}

\noindent where $a_{\mathrm{sky}}$ is the upper angular acceleration constraint of Sgr A*-IR on the sky over its observed time baseline, $R_{\mathrm{GC}}$ is the distance to the GC, and $G$ is the gravitational constant. It is assumed that $M_{\mathrm{IMBH}} \ll M_{\mathrm{SMBH}}$.

For the second bound, the astrometric uncertainty determines the maximum allowable size of the astrometric wobble of Sgr A* due to an IMBH companion. Below our astrometric precision, we would mistake any physically induced ``wobble" of the Sgr A*-IR  position as noise in our measurements. This bound represents the point at which we cannot tell the difference between actual wobble of Sgr A*-IR and error in our measurements:

\begin{equation}
    M_{\mathrm{IMBH}} < M_{\mathrm{SMBH}}\left(\frac{R_{\mathrm{GC}}}{r}\right) \delta\theta
\end{equation}

\noindent where $\delta\theta$ is the average astrometric error on our model fit to Sgr A*-IR ($\delta\theta = 2.07$ mas) and $M_{\mathrm{SMBH}}$ is the mass of the SMBH associated with Sgr A*. The intersection point of the above-stated IMBH constraints limit us to a $\lesssim 4 \times 10^{4}$ $M_{\odot}$ IMBH at a semi-major orbital axis of $\sim 0.01$ pc.

\begin{figure}
    \centering
    \includegraphics[width=1\linewidth]{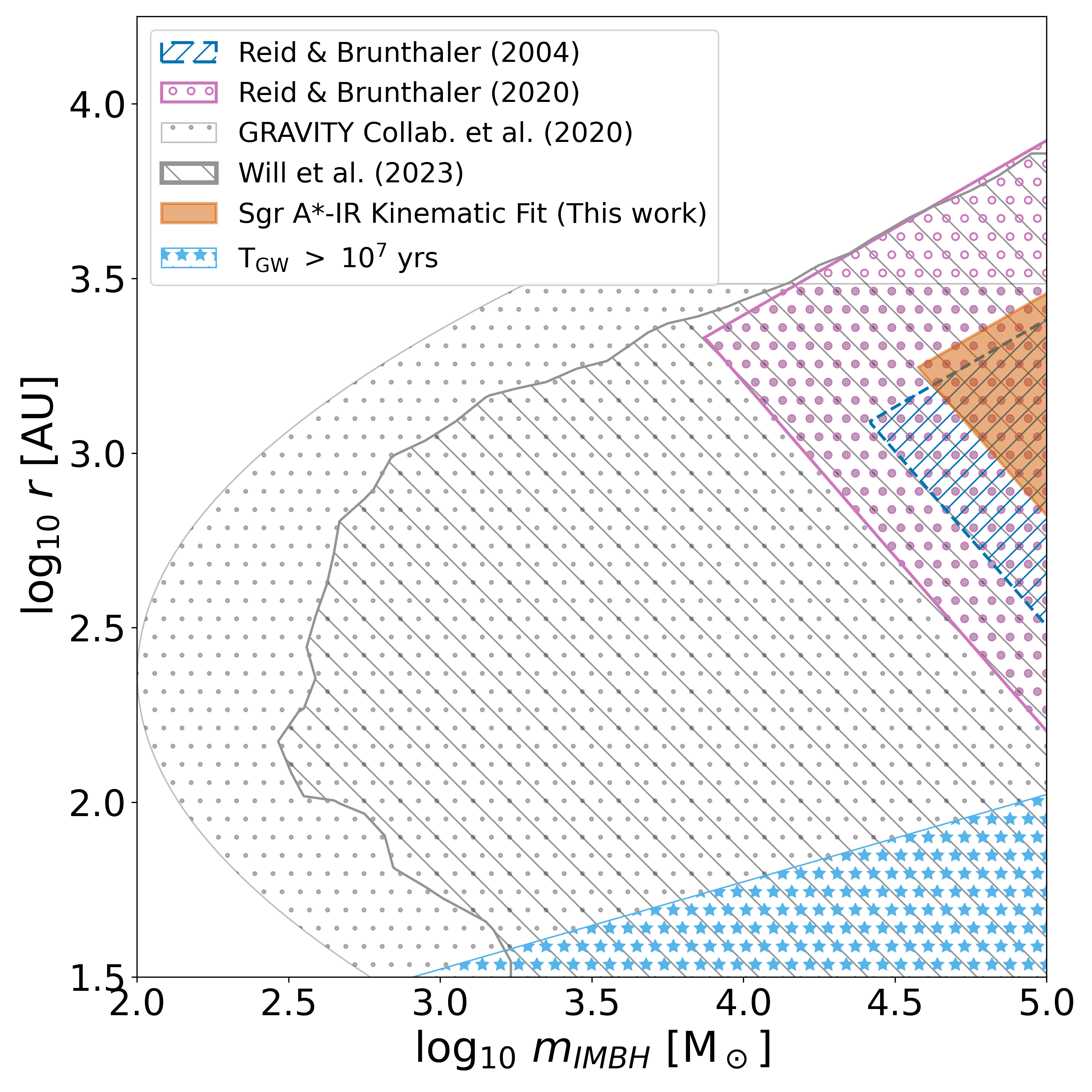}
    \caption{Excluded regions of parameter space for an IMBH companion to the SMBH associated with Sgr A*. The excluded parameter space by this work is shaded in solid orange and is bounded by the upper limit on the acceleration measured for Sgr A*-IR and the average precision of the Sgr A*-IR astrometry. Similar bounds are used for the excluded zones for \citet{Reid_2004, Reid_2020}, shown as shaded with blue left-to-right hatches and with pink circles, respectively. \citet[][shaded in grey right-to-left hatches]{Will_2023} and \citet[][gray dots]{Abuter_2020} use changes in the orbital parameters of S0-2 over time to constrain a potential IMBH companion. Plotted in cyan stars is the region that would exclude an IMBH companion because within a time of $>10^{7}$ yr, the IMBH would merge into the SMBH \citep{Naoz_2020, Will_2023}.}
    \label{fig:imbh_params}
\end{figure}

We show our constraints in context with previous literature in Figure \ref{fig:imbh_params}. All shaded regions on the plot are masses or orbital semi-major axes that are excluded from the parameter space \citep{Reid_2004, Reid_2020, Abuter_2020, Naoz_2020, Will_2023}. Our current constraints fall well within the constraints determined by other works, but this work represents the first limits derived from the absolute motion of Sgr A*-IR. This exercise shows that acceleration limits on Sgr A*-IR may be used to constrain parameters on a hypothetical IMBH companion. The precision of our constraints is expected to improve in future work (Section \ref{subsec:future_improvs}). 

\subsection{Expected Future Improvements with Gaia Data Release 4}
\label{subsec:future_improvs}

The Gaia mission plans to release a Data Release 4 (DR4) astrometric catalog sometime in 2026 based on 66 months of data taken by the Gaia spacecraft. \footnote{The Gaia mission data release schedule is found at https://www.cosmos.esa.int/web/gaia/release.} The HST-Gaia catalog created by \citet{Hosek_2025} provides the reference stars that we use to transform our Keck AO star lists into the Gaia-CRF3 reference frame. In a similar framework, Gaia DR3 provided the reference stars that \citet{Hosek_2025} used to transform their HST observations into the Gaia-CRF3 reference frame in order to create the HST-Gaia catalog. If Gaia DR4 is used to recreate the HST-Gaia catalog, providing improved measurements and uncertainties on the reference stars, then Gaia DR4 is expected to improve the astrometric uncertainties of stars in the HST-Gaia catalog by 1.6 times and to improve their proper motion uncertainties by 2.4 times \citep{Hosek_2025}. Therefore, if we redo the transformation of our Keck AO data into the Gaia Celestial Reference Frame (Gaia-CRF), we would expect a DR4-based HST-Gaia catalog to provide astrometric reference stars with improved  measurements. These DR4-based HST-Gaia catalog reference stars should yield lower transformation uncertainties for the Keck AO transformation into Gaia-CRF. 

Per observation epoch, \citet{Hosek_2025} estimated the expected transformation uncertainty in Gaia-CRF3 as

\begin{equation}
    \sigma_{\mathrm{trans}} = \alpha_{\mathrm{obs}}\frac{\overline{\sigma_{\mathrm{pos}}}}{\sqrt{N_{\mathrm{ref} - n_{\mathrm{params}}}}}
\end{equation}

\noindent where $\overline{\sigma_{\mathrm{pos}}}$ is the median total positional uncertainty of the reference stars in that epoch, $N_{\mathrm{ref}}$ is the number of reference stars, $n_{\mathrm{params}}$ is the number of parameters in the transformation (six for the second-order polynomials used in this work), and $\alpha_{\mathrm{obs}}$ is an empirically determined constant. \citet{Hosek_2025} had calibrated $\alpha_{\mathrm{obs}}$ for the transformation of stars in their HST star lists into Gaia-CRF3 using Gaia DR3 as the source of reference stars. They had found that $\alpha_{\mathrm{obs}} = 2.6$, allowing them to calculate to the previously mentioned expected improvements in uncertainties if Gaia DR4 is used.

We repeat this procedure to determine our expected improvements for our Keck AO star lists using an improved HST-Gaia catalog based on Gaia DR4. For each reference star in this work, $\overline{\sigma_{\mathrm{pos}}}$ is the quadratic sum of its HST-Gaia catalog uncertainty and its total astrometric error $\sigma_{\mathrm{total}}$ in the Keck AO star lists. We also assume here that the number of reference stars from the DR4-based HST-Gaia catalog is the same as we currently use, and as such, our improvement estimate here is conservative. We would expect further improvements if additional reference stars in Gaia DR4 meet our astrometric quality criteria. With this, for the transformation errors of the Keck AO star lists into Gaia-CRF3, we find that $\alpha_{\mathrm{obs}} = 4.1$ in R.A.$\cos{\mathrm{DEC}}$ and $\alpha_{\mathrm{obs}} = 3.2$ in DEC. In addition, as described in Section \ref{subsec:trans_keck_to_gaia}, the transformation is directly affected by the quality of proper motion measurements of the reference stars used to calculate it. Any improvements in the proper motion measurements in the transformed Keck AO star lists would result in improvements in the reference stars in the HST-Gaia catalog, and so we take these expected proper motion improvements to be the same as for the HST-Gaia catalog in \citet{Hosek_2025}.

Altogether, if Gaia DR4 were used to recreate the HST-Gaia catalog, we would expect the transformation uncertainties of our reference stars in the transformed Keck AO star lists to be between 0.21 and 0.37 mas in R.A.$\cos{\mathrm{DEC}}$ and between 0.16 and 0.28 mas in DEC. This would represent a 1.3 times improvement in R.A.$\cos{\mathrm{DEC}}$ and a 1.8 times improvement in DEC for transformation errors in the Keck AO star lists. Also of importance are how these increases in precision would affect the proper motion uncertainties. In the HST-Gaia catalog, the proper motion uncertainty of a given star approximately scales with the average transformation error $\sigma_{\mathrm{trans}}$ across epochs such that $\sigma_{\mathrm{pm}} \propto \sigma_{\mathrm{trans}}$.  With this we would expect an improvement in the Keck AO star list proper motion uncertainties of 1.4 times in R.A.$\cos{\mathrm{DEC}}$ and 2.1 times in DEC. This estimate is conservative with our assumption of the same number of reference stars. 

We would expect these increases in astrometric and proper motion precision to propagate to our measurements of the kinematic motion of Sgr A*-IR. However, for Sgr A*-IR, the positional uncertainties ($\sigma_{\mathrm{pos}}$ as defined in Section \ref{subsec:keck_ao_obs}) currently dominate over our transformation uncertainties. The largest improvement we would need to make to this work in order to obtain a better kinematic fit to Sgr A*-IR is to better model the scatter in points that currently require a constant additive systematic error to compensate. If the systematic uncertainty model in the kinematic fit to Sgr A*-IR is improved, the improvements we expect in astrometric precision with Gaia DR4 would improve our kinematic fit and our IMBH parameter constraints by a factor of about 2.

\section{Conclusion}
\label{sec:conclusion}

We use 17 yr of Keck AO data in \textit{K}-band to measure the absolute motion of Sgr A*-IR for the first time. From the Keck AO observations, we retrieved 14 epochs of Sgr A*-IR positions that are not confused with known stars out of 44 epochs of observation. Previously, an absolute motion measurement of Sgr A*-IR was not possible due to the lack of an absolute IR reference frame for the central 10" of the GC. Thus, we transformed our Keck AO observations into the Gaia-CRF3 reference frame, which is defined relative to distant quasars and thus is an absolute rather than relative reference frame. We used 32 reference stars from the HST-Gaia catalog produced by \citet{Hosek_2025} to transform our Keck AO observations into Gaia-CRF3. Using bright stars in the Keck AO star lists to evaluate the quality of our transformation, we achieved median transformation uncertainties between 0.14 to 0.66 mas across our 44 epochs. When we evaluate how well we transform the reference stars themselves into Gaia-CRF3, we find that the proper motion bias between the HST-Gaia catalog and the transformed Keck AO star lists is consistent with zero and within a precision of $\sim 0.01$ mas yr$^{-1}$. However, the position bias of the reference stars does show a nonzero offset of 0.550 $\pm$ 0.055 mas in R.A.$\cos{\mathrm{DEC}}$ and -0.366 $\pm$ 0.066 mas in DEC compared to the HST-Gaia catalog.

We fit a kinematic model to the Sgr A*-IR astrometry using Gaussian processes in the same methodology as \citet{Hosek_2025}. The kinematic models used are a combination of a first- or second-order polynomial and a systematic uncertainty model. Both models are fit to the data simultaneously. We find that the best-fit model to the Sgr A*-IR data is a first-order polynomial with a constant additive error to account for systematic uncertainties. The best-fit proper motion is $\mu_{\alpha^{*}} = -3.093 \pm 0.085$ mas yr$^{-1}$ and $\mu_{\delta} = -5.62 \pm 0.13$ mas yr$^{-1}$. The $t = 2016.0$ position is ($\alpha(t)$, $\delta(t)$) = (266.41680848 $\pm$ 0.00000029, -29.00783947 $\pm$ 0.00000050) deg. The constant additive errors incorporated into this fit to account for systematic uncertainty are $\sigma_{\mathrm{add}, \alpha^{*}} = 1.05$ mas and $\sigma_{\mathrm{add}, \delta} = 1.79$ mas, likely caused by confusion with unresolved sources near Sgr A*. Our kinematic motion model fit for Sgr A*-IR is consistent with the measured motion of Sgr A*-radio as found by \citet{Xu_2022}. In addition, our second-order polynomial motion model fit to Sgr A*-IR, while not the most preferred model, does allow us to put constraints on the acceleration on the sky of Sgr A*-IR. The fitted acceleration is consistent with zero and we put a $2\sigma$ upper limit on the acceleration of $0.061$ mas yr$^{-2}$. 

Between our acceleration upper limit on Sgr A*-IR and our average astrometric uncertainty on the model fit, we can also put constraints on the parameter space that could be occupied by a hypothetical IMBH companion to the SMBH associated with Sgr A*. Overall, our IMBH constraints limit us to a $\lesssim 4 \times 10^{4}$ $M\odot$ IMBH at a semi-major orbital radius of $\lesssim 0.01$ pc. This is the first constraint put on the reflex motion of the SMBH associated with Sgr A* using the absolute motion of its IR counterpart.

Moving forward, we expect that Gaia DR4 will improve the astrometric precision of the HST-Gaia catalog stars used to transform the Keck AO star lists into Gaia-CRF. However, the largest source of uncertainty for the Sgr A*-IR motion fit is currently the scatter in the astrometry, which requires a constant additive error to account for. Further improvements in our handling of unknown source confusion and astrometric biasing with Sgr A*-IR will be needed before we can see the improvements we expect from a Gaia DR4-based HST-Gaia catalog to propagate to our kinematic motion fit of Sgr A*-IR.  Assuming the improvements we expect in the astrometric precision with Gaia DR4, our IMBH constraints would limit the parameter space to $\sim2\times10^{4} $ $M_{\odot}$ at \textit{R}$<$0.01 pc, improving our constraints by a factor of at least 2, and further improvement is expected with increased numbers of quality reference stars. Altogether, Sgr A*-IR provides a significant astrometric reference in studying the environment around its associated SMBH. 

\begin{acknowledgments}

M.W.H. is supported by the Brinson Prize Fellowship. A.M.G. acknowledges support from her Lauren B. Leichtman and Arthur E. Levine Endowed Astronomy Chair. The authors acknowledge support from the Gordon E. \& Betty I. Moore Foundation (award \#11458) and GC Star Society. This work is based on observations made with the NASA/ESA Hubble Space Telescope, obtained at the Space Telescope Science Institute, which is operated by the Association of Universities for Research in Astronomy, Inc., under NASA contract NAS 5-26555. The observations are associated with programs GO-11671, GO12318, GO-12667, GO-13049, GO-15199, GO-15498, GO-16004, GO-15894, GO-16681, and GO-16990. This research has made extensive use of the NASA Astrophysical Data System.

\textit{Software}: AstroPy \citep{Astropy_2022}, Matplotlib \citep{Hunter_2007}, NumPy \citep{Harris_2020}, SciPy \citep{Virtanen_2020}

\end{acknowledgments}

\appendix

\section{Assessing transformation errors into Gaia-CRF3}
\label{appendix:astrometric_assessment}

We assess the estimation of total astrometric error $\sigma_{\mathrm{total}}$ for stars in the transformed Keck AO observations using the reduced chi-squared ($\chi^{2}_{\mathrm{red}}$) distribution of their proper motion fits. For this, we crossmatch our Keck AO star lists against a catalog of ``good stars" put together by \citet{Jia_2019}. These ``good stars" were used by \citet{Jia_2019} to determine an optimal additive error function to accommodate for astrometric error not covered by $\sigma_{\mathrm{cent}}$ or $\sigma_{\mathrm{trans}}$. We were able to crossmatch 341 of 352 sources from the ``good stars" catalog. 

\begin{figure}
    \includegraphics[width=1\linewidth]{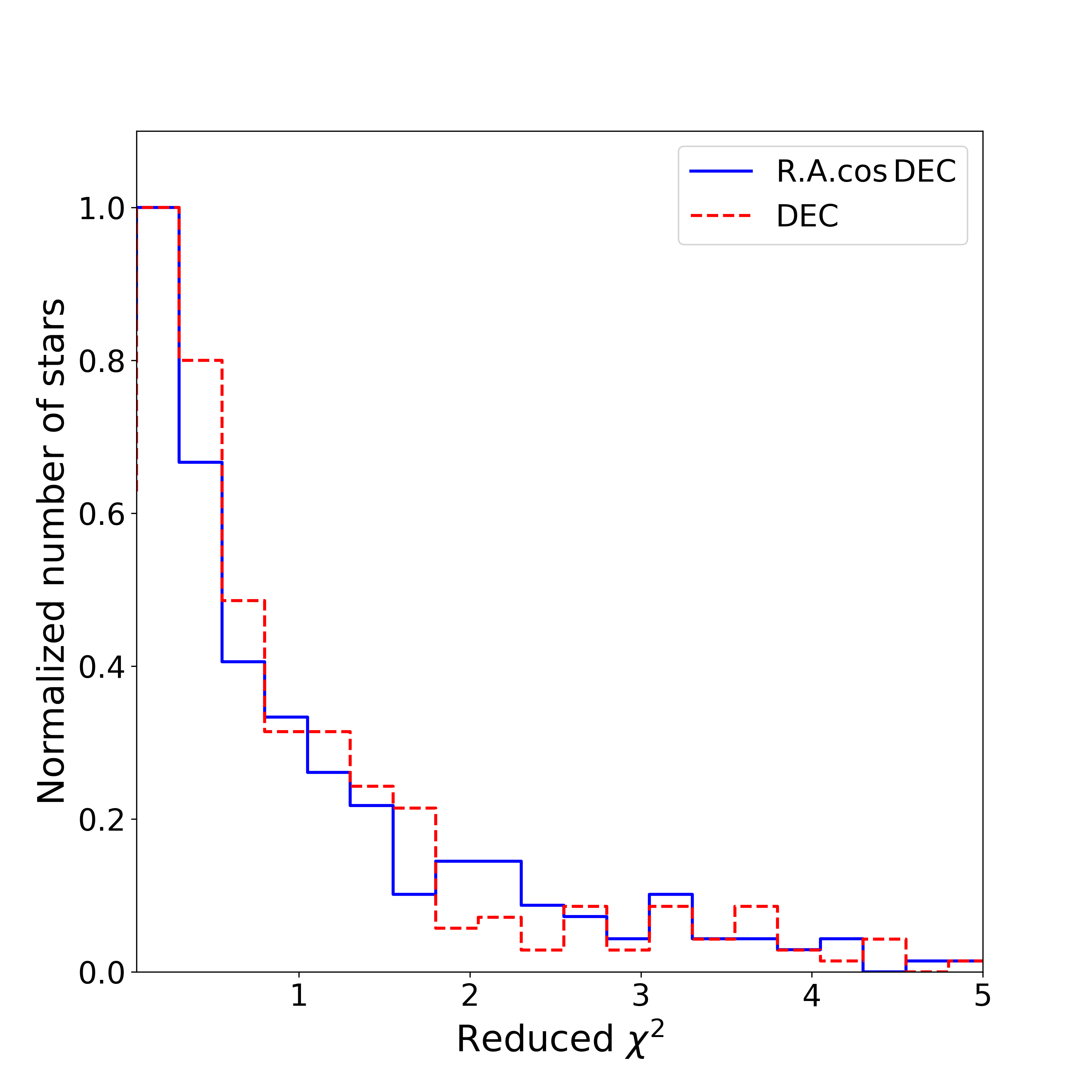}
    \caption{Reduced $\chi^{2}$ PDF distributions of proper motion fits to the ``good stars" \citep{Jia_2019} in our transformation to Gaia-CRF3. The observed PDF for R.A.$\cos{\mathrm{DEC}}$ is shown as a solid blue line and the one for DEC is shown as a dashed red line.}
    \label{fig:red_chi2}
\end{figure}

We compare the Keck AO stars' posttransformation $\chi_{\mathrm{red}}^{2}$ probability density function (PDF) to a theoretical $\chi_{\mathrm{red}}^{2}$, where $\chi_{\mathrm{red}}^{2}$ is calculated as

\begin{equation}
    \chi_{\mathrm{red}}^{2} = \sum_{i = 1}^{N} \frac{1}{N-2}\left( \frac{p_{\mathrm{obs},i} - p_{\mathrm{fit},i}}{\sigma_{\mathrm{total},i}}\right)
\end{equation}

\noindent such that a given star has $N$ observed points, $p_{\mathrm{obs},i}$ is the observed position of the star at point $i$, $p_{\mathrm{fit},i}$ is the expected position of the star from its proper motion fit, and $\sigma_{\mathrm{total},i}$ is the total astrometric uncertainty of each observed position. Underestimated astrometric error on average would lead the observed PDF to peak at values greater than $\chi^{2}_{\mathrm{red}}$ = 1.0. Overestimated error on average would lead the observed PDF to peak at values less than $\chi^{2}_{\mathrm{red}}$ = 1.0. We find that the PDF peaks well below the $\chi^{2}_{\mathrm{red}}$ = 1.0, indicating that the uncertainties on average are overestimated, as shown in Figure \ref{fig:red_chi2}.

The Keck AO stars' astrometric errors are dominated by transformation errors, as is shown in Figure \ref{fig:final_err_vs_epoch}. We believe that may be due to the method used to determine the transformation uncertainties. Following the transformation procedure of \citet{Jia_2019}, the transformation errors are calculated by performing $N=100$ half-sample bootstraps over the reference stars and calculating the transformation for each one-half of the sample. The transformation uncertainty of a given star is the standard deviation of the 100 transformed positions. However, because our selection of reference stars contains only 32 stars, this results in the half-sample transformations using $\sim$16 stars each. This may not be enough to perform a quality transformation, resulting in a wide spread of transformed positions and artificially increasing the transformation error. Thus, the transformation error of the stars may be overestimated, and so we consider it a conservative estimate of the transformation error of the Keck AO stars.

\section{Expected Log Predictive Density}
\label{appendix:elpds}

Here, we review the use of ELPD for model selection for this work. For more general cases, see \citet{Gelman_2013} and \citet{Vehtari2017}. We consider here the specific case of evaluating leave-one-out residuals and ELPDs for the kinematic models with and without accelerations using the Gaussian processes kinematic model from \citet{Hosek_2025}. 

Consider a source that is observed to have positions $\mathbf{x_{obs}} = \{x_{\mathrm{obs},i}\}$ that are generated by a true model $f$. This true model $f$ is unknown. Let $\mathbf{x_{\ast}} = \{x_{\ast,i}\}$ be measurements generated by model $f$ that are not part of the set $\mathbf{x_{obs}}$. The data distribution of $\mathbf{x_{\ast}}$ generated by model $f$ is $f(\mathbf{x_{\ast}})$. Additionally, let the model we are fitting to $\mathbf{x_{obs}}$ be model $k$ with parameters $\theta_{k}$. 

With these terms, the ELPD of model $k$ can be written as

\begin{equation}
\label{eqn:ELPD}
    \mathrm{ELPD}^{(k)} \equiv \int f(x_{\ast,i}) \, \mathrm{ln}(p_{k}(x_{\ast,i} | \mathbf{x_{obs}})) \, dx_{\ast,i}
\end{equation}

\noindent where $p_{k}(x_{\ast,i} | \mathbf{x_{obs}}) = \int p_{k}(x_{\ast,i} | \theta_{k}) \, p_{k}(\theta_{k} | \mathbf{x_{obs}}) d\theta_{k}$ is the probability density of observing $x_{\ast,i}$ given model $k$.

Because $f(x_{\ast,i})$ is unknown, we must approximate the ELPD. We used leave-one-out cross-validation (LOO-CV) because it has been shown to asymptotically approximate the ELPD \citep{Watanabe_2010}. In LOO-CV, we fit model $k$ on the observed dataset $\{x_{obs,j}\}_{j \neq i}$ where one data point $x_{obs,i}$ is excluded. The probability density of observing $x_{obs,i}$ given $\{x_{obs,j}\}_{j \neq i}$ is 

\begin{eqnarray}
        p_{k}(x_{obs,i} | \{x_{obs,j}\}_{j \neq i}) = \int p_{k}(x_{obs,i} | \{x_{obs,j}\}_{j \neq i}, \theta_{k}) \nonumber \\
        p_{k}(\theta_{k} | \{x_{obs,j}\}_{j \neq i}) d\theta_{k} \nonumber \\
\end{eqnarray}

Therefore, the approximated ELPD using LOO-CV is

\begin{equation}
    \mathrm{ELPD}^{(k)} \approx \widehat{\mathrm{ELPD}}\mathrm{^{(k)}_{loo-cv}} \equiv \sum_{i} \mathrm{ln}(p_{k}(x_{obs,i} | \{x_{obs,j}\}_{j \neq i}))
\end{equation}

Because we are assuming Gaussian processes, this is calculated as

\begin{equation}
    \widehat{\mathrm{ELPD}}\mathrm{^{(k)}_{loo-cv}} \propto -\frac{1}{2}\sum_i \frac{(x_{obs, i} - x_{pred, i})^2}{\sigma^2_{pred,i}} + \mathrm{log}(\sigma^2_{pred, i})
\end{equation}

\noindent where $x_{pred, i}$, and $\sigma_{pred, i}$ are, respectively, the prediction and uncertainty of the data point $i$ where that prediction is done by an analysis that excludes the $x_{obs, i}$ data point.

Now, for the given source, let model $a$ be the Gaussian processes model with only a polynomial kernel and no systematic uncertainty kernel (single-component model). Let models $b$, $c$, and $d$ be the polynomial kernel plus the squared-exponential kernel, the confusion kernel, and the constant additive error, respectively. We need to select which model ($a$, $b$, $c$, or $d$) best fits the observed positions of the source. The simplest model here is model $a$ as it includes no systematic uncertainty kernel, and so, if models $b$, $c$, and $d$ do not have a better ELPD than model $a$, model $a$ is chosen as the best-fit model. 

To compare the ELPDs, we take the difference of model $i$ where $i \in b,c,$ and $d$ and model $a$:

\begin{equation}
    \Delta \widehat{\mathrm{ELPD}}_{\mathrm{{loo-cv}}}  = \widehat{\mathrm{ELPD}}_{\mathrm{{loo-cv}}}^{(i)} - \widehat{\mathrm{ELPD}}_{\mathrm{{loo-cv}}}^{(a)}
\end{equation}

\noindent since this is equivalent to a ratio of the models' ELPDs.

To select model $i$ over model $a$ as the best fit for a source, we determined that model $i$ needed to be preferred by a probability threshold of $99.7\%$ ($3\sigma$) over model $a$, meaning that the probability of the simpler model being favored must be less than $0.003$. This corresponds to the requirement that $\Delta \widehat{\mathrm{ELPD}}\mathrm{_{loo-cv}}$ must be $\gtrapprox 6$. 

Therefore, if  $\Delta \widehat{\mathrm{ELPD}}\mathrm{_{loo-cv}} \geq 6$, then the model $i$ is selected over model $a$ and if multiple models $i$ satisfy the condition $\Delta \widehat{\mathrm{ELPD}}\mathrm{_{loo-cv}}  \geq 6$, then the model $i$ with the largest $\Delta \widehat{\mathrm{ELPD}}\mathrm{_{loo-cv}}$ is selected.

\bibliography{HSTGaiaGClimits}
\bibliographystyle{aasjournal}

\end{document}